\documentclass[prd,twocolumn,preprintnumbers,superscriptaddress,nofootinbib]{revtex4}
\usepackage[a4paper, hdivide={1.91cm,,1.165cm}, vdivide={1.83cm,,2.6cm}]{geometry}

\usepackage{amstext,amssymb}
\usepackage{amsmath,lipsum}
\usepackage{graphicx}
\usepackage{subcaption}
\usepackage{color}
\usepackage[toc,page]{appendix}
\usepackage[hyperfootnotes=false]{hyperref}

    \usepackage{hyperref}
\hypersetup{ 
	setpagesize=false,
	bookmarksnumbered=true,
	colorlinks=true,
	linkcolor=blue,
	citecolor=red,
	hypertexnames=true
}
\usepackage{xspace}
\usepackage{color}
\usepackage{caption}
\usepackage{subcaption}
\usepackage{float}
\usepackage{units}
\usepackage{slashed} 
\usepackage{braket}

\def \vec#1{{\boldsymbol{#1}}}

\begin{document}

\title{\hspace*{-1.0cm} Imprints of LGI violation in Mesons}


\author{Kiran \surname{Sharma}}
 \email{kirans@iitbhilai.ac.in}
\affiliation{Department of Physics, Indian Institute of Technology Bhilai, India}
\author{Aryabrat \surname{Mahapatra}}
 \email{21ph05015@iitbbs.ac.in}
\affiliation{School of Basic Sciences, Indian Institute of Technology Bhubaneswar, India}
\author{Prasanta K. \surname{Panigrahi}}
 \email{panigrahi.iiser@gmail.com}
\affiliation{Indian Institute of Science Education and Research Kolkata, 741246, WB, India}
\author{Sudhanwa \surname{Patra}}
 \email{sudhanwa@iitbhilai.ac.in}
\affiliation{Department of Physics, Indian Institute of Technology Bhilai, India}

\begin{abstract}
Quantum mechanics has always proven emphatically as one of the main cornerstones in all of science since its inception. Initially, it has also faced many skeptics from many scientific proponents of its complete description of reality. However, John Bell once devised a theorem on quantitative grounds to show how local realism is expressed in quantum mechanics. As Bell's inequality claims the non-existence of local hidden variable theories, on the same footing, we have Leggett-Garg's Inequality (LGI), which sets a quantum-classical limit for temporally correlated quantum systems. In the context of particle physics, specially in the field of neutrino and meson oscillations, one can conveniently implement LGI to test the quantum foundations at the probability level. Here, we discuss the significant LGI violation characteristics in B- and K- meson oscillations, taking into account their decoherence, CP violation, and decay parameters. We have emphasized the fact that \textit{Tsirelson} bounds can be achieved under certain conditions. Our focal point is to show the signature of these bounds that appears only to specific alterations of decay and decoherence effects. Also, we discuss and comment on the behavior or effect of decoherence and decay widths playing out from the perspective of LGI by taking their available values from various experiments. This may help us to understand the underlying principles and techniques of neutral meson open quantum systems.

\vspace*{-1.0cm}
\end{abstract}


\maketitle

\section{Introduction}
\noindent

Right from the advent of quantum mechanics, we saw appealing success in all of its predictions and found flourishing consistent results based on real-life applications. However, some scientific proponents
questioned the very core ideas and principles of quantum mechanics. Among them, the most notable strike came in 1935, when Albert Einstein, Boris Podolsky and Nathan Rosen published the famous EPR paradox~\cite{Einstein:1935rr} which doubts the indeterministic framework and formalism of quantum mechanics. The fundamental assumption on which the EPR argument rests is the principle of locality. They argued the interpretation of Copenhagen formalism is incomplete where wave functions instantaneously collapse upon apparently non-local measurement. This contradicts the viewpoint of quantum mechanics to have a definite value of an observable prior to measurement. The more appropriate version of this idea was actually proposed by D. Bohm through EPR-Bohm experimental model~\cite{Bohm:1957zz}, where he considered the system of two atoms/particles (like an $e^- e^+$ pair) each with spin half prepared with the total spin zero (in singlet state configuration). This Bohm model is one of the first explicitly non-local hidden variable models. 
\newline
In 1964, John Bell visited this EPR experiment and devised a quantitative mechanism through inequalities to test these hidden variable theories. This inequality was deduced under the assumption oriented with EPR claim that there was some hidden variable underlying quantum mechanics. Also, this inequality has been well verified over several occasions and the first experimental verification was carried out by Aspect and co. in 1980s (the latest Nobel laureates for this work). It turned out that this inequality is violated, and thus it is incompatible with quantum mechanical interpretations. The idea of two particles separated by an arbitrarily large range has to be superluminal, and hence it is non-local. These Bell inequalities turn the EPR argument completely on its head and forbid any local hidden variable theory and indeed, it proves that any correlations that arise in quantum mechanics cannot arise in any classical or macroscopic system.

In the same outline, Leggett and Garg introduced Leggett-Garg Inequality \cite{Leggett:1985zz} which attempts to test quantum foundations in various correlated quantum systems by considering certain classical assumptions. This mathematical inequality fulfills the macroscopic physical theories, and explains the quantum behavior upon its violation just like Bell inequality. We will implement this technique in open quantum systems \cite{BRE02} of the neutral meson-antimeson oscillatory system, a two-state decaying system that can act as a bipartite entangled system.

In this work, we have set up the measurement inequalities from the LGI formalism. Then, we have given a formal introduction to the fundamental description of meson oscillations and their respective probabilities, where we have assumed that CP remains conserved in the entire propagation states of mesons. Although, there has been a lot of successful work done on this subject \cite{ Banerjee:2014vga, Naikoo:2019gme,Naikoo:2018vug,PhysRevLett.113.050401, Naikoo:2019eec, Naikoo:2017fos,Dixit:2018gjc,Gangopadhyay:2013aha,PhysRevA.91.032117, Shafaq:2021lju,S.Banerjee,Shafaq:2020sqo,PRXQuantum.3.010307}; however, we have highlighted certain imprints of LGI violations that attempt to bring out some specific behavior of LGI function, which is perhaps unique and distinct from other works. Our work has some similarities with work in~\cite{Naikoo:2018vug} where authors looked at the variation of 3-measurement LGI parameter $K_3$ with various time intervals $\Delta t$ for $K_s$ and $B_d$ mesons. Both papers have considered decay and decoherence effects together and found that, without any surprise, the results coincide. But there are some significant differences in both works, which are described as follows: 
\newline
We have highlighted the feature when decay and decoherence effects are studied individually; then, it is observed that the system allows a distinct ability to reach saturation, the '\textit{Tsirelson bound}'. This suggests the fact that decay and decoherence time have different physical origins because our observations will indicate that the possibility of reaching the quantum bound is distinct for each of these effects. To enable comparison with physical decay and decoherence times, we plot with the absolute or direct time interval $\Delta t$. Also, in this work, naive LGI testing is carried out for CP-conserved eigenstates of mesons which shows a strange result of the over-enhanced violation of LGI. The primary takeaway from this work is to understand how and in what ways this \textit{Tsirelson bound} can be obtained in meson oscillatory system and realize them with reasonable interpretation.

\section{Theoretical framework}
In this section, we will resort to the understanding of the theoretical background of LGI and meson oscillations. We generalize a methodology to define a correlation function in LGI in terms of macroscopic variables which can only take two values. This mechanism has wide applicability in testing LGI violation from photon systems \cite{Goggin}  to diamond defect centers, NMR systems \cite{Katiyar_2017}, superconducting flux qubits~\cite{Nature}, solar neutrinos \cite{Murskyj, Shafaq:2021lju,Shafaq:2020sqo} and more. Then, we will discuss the phenomenology of the two-state oscillatory system of K- and B-mesons which are generally produced in strong and electromagnetic interactions and they decay by weak interactions. Finally, we will use the probabilistic approach of decaying systems \cite{PhysRevA.31.1736, Banerjee:2014vga,PhysRevA.104.052225} to study various measures of quantum correlations in $B\Bar{B}$ and $K\Bar{K}$ oscillation systems.

\subsection{LGI Formalism:}
The sole motive for proposing LGI is to understand how macroscopic realism accounts for quantum systems and how the idea of certainties and superposition of the classical world can govern the principles in the microscopic domain. The assumptions on which the entire paradigm of LGI rests are macroscopic realism and noninvasive measurability \cite{Emary_2014,PhysRevA.91.062103,Conference}. According to Leggett and Garg, macroscopic realism is the worldview in which attributes of macroscopic systems exist independent of external measurement and observation.  And noninvasive measurability says that a result of observation is not influenced anyway by the mere act of measurement unlike what Copenhagen interpretation demands. Classical physics works very well relying on these assumptions, while quantum mechanics is downright inconsistent. 

\subsubsection{Setting up LGI in terms of correlation functions:}
Suppose, if we consider a macroscopic variable that can take on two values, $Q=\pm1$, and its two-time correlation function  \cite{Murskyj}
\begin{eqnarray}
 C_{ij}=\langle \hat{Q}(t_i)\hat{Q}(t_j) \rangle
\end{eqnarray}
  where  $\hat{Q}(t)$ is the operator corresponding to the observable Q.If macroscopic realism
and noninvasive measurability are obeyed, then the Leggett-Garg inequality places
classical limits on a combination of the correlations: for three measurements,
\begin{eqnarray}
 -3\le C_{12}+C_{23}-C_{13}\le 1
\end{eqnarray}

In general, there can be LGIs corresponding to an n-measurement Leggett-Garg parameter,
\begin{eqnarray}
 -n \le K_n\le n-2, \hspace{0.29cm} n=3,5,7,.. \nonumber\\
 -(n-2) \le K_n\le n-2, \hspace{0.29cm} n=4,5,6,.. 
\end{eqnarray}

where, $K_n=C_{12}+...+C_{n-1\hspace{0.05cm}n}-C_{n1}$
We can prove these inequalities in the following way:\\
The correlation function $C_{ij}$ is defined as;\\
\begin{eqnarray}\label{eqn-4}
 C_{ij}=\sum_{Q_i, Q_j=\pm1}Q_iQ_jP_{ij}(Q_i,Q_j)
\end{eqnarray}
where $P_{i,j}(Q_i,Q_j)$ is the joint probability of measuring $Q_i=\pm1$ at time $t_i$ and $Q_j=\pm1$
at time $t_j$\\
Now, we can express the above joint probability in terms of some third possible measurement observable at a different time taking these same values;
\begin{eqnarray}\label{eqn-5}
 P_{ij}(Q_i,Q_j)=\sum_{Q_k=\pm1,k\ne i,j}P_{ij}(Q_1,Q_2, Q_3)
\end{eqnarray}
where  $Q_3$ is the value of Q at $t=t_3$.

So, we know $K_3=C_{12}+C_{23}-C_{13}$ and hence $K_3$ can be written as

\begin{eqnarray}\label{eqn-6}
 K_3=\sum_{Q_i, Q_j=\pm1} (\sum_{Q_k, k\ne 1,2} P(Q_1,Q_2, Q_3)+\nonumber\\
 \sum_{Q_k, k\ne 2,3} P(Q_1,Q_2, Q_3)-\sum_{Q_k, k\ne 1,3} P(Q_1,Q_2, Q_3) )
\end{eqnarray}

Now, using the fact of normalization of probabilities, we can write,  $\Sigma=1$. So, then $K_3$ will reduce to the equation;
\begin{eqnarray}
 K_3=1-4\left( P(+,-,+)+P(-,+,-) \right)
\end{eqnarray}

Then the minimum and maximum bounds for $K_3$ will go as:
\begin{eqnarray}\label{eqn-8}
 -3\le C_{12}+C_{23}-C_{13}\le 1
\end{eqnarray}
Finally, equation (\ref{eqn-8}) gives the so-called Leggett-Garg Inequality for a three-measurement correlation.

\subsubsection{LGI Violation}

This correlation function has been laid out based on classical assumptions, so they can be well treated as classical physical observables. If so, then classical observables commute with each other and follow the symmetric relation among them. Thus, the correlation function $C_{ij}=\langle Q_iQ_j \rangle$ can be given as:
\begin{eqnarray}
 C_{ij}=\frac{1}{2}\langle Q_iQ_j + Q_jQ_i \rangle 
\end{eqnarray}

Suppose, we view these parameters as quantum mechanical observables. It can be possible to express them on the basis of the linear combination of Pauli matrices i.e., $\hat{Q_i}=\vec{a_i}.\hat{\sigma}$, then we find that
\begin{eqnarray}
 &C_{ij} = \frac{1}{2}\langle Q_iQ_j + Q_jQ_i \rangle = \vec{a_i}.\vec{a_j}&\\
 &\left( \because (\vec{\sigma}.\vec{A})(\vec{\sigma}.\vec{B})=\vec{A}.\vec{B}+i\vec{\sigma}.(\vec{A}\times\vec{B})\right)&\nonumber\\
 \implies &K_3 = \cos{\theta_1}+\cos{\theta_2}-\cos{(\theta_1+\theta_2)}&
\end{eqnarray}

where, $\theta_1$ is the angle between vectors $\vec{a_1}$ and $\vec{a_2}$, $\theta_2$ is between vectors $\vec{a_2}$ and $\vec{a_3}$.

Now, to get the upper limit of $K_3$, we have to choose a specific set of values for $\theta s$. We see when $\theta_1=\theta_2=\theta_3=\frac{\pi}{3}$ we get the maximum value of $K_3$, $max(K_3)=1.5$

This is beyond the predicted value of LGI; hence, these quantum mechanical observable indeed violates the inequality.

Now, we can draw an analogy between LGI and Bell Inequality. Looking at quantum mechanics unlike classical observables we find, 
\begin{eqnarray}
 \left[ Q_i, Q_j\right]=2i\hat{\sigma}.(\vec{a_i}\times \vec{a_j})
\end{eqnarray}
instead of zero. Thus, it violates LGI in the quantum regime.
Bell inequality sets a lower bound and an upper bound for spatially correlated events. Similarly, LGI sets the limits for temporally correlated events. Hence, LGIs are also known as temporal Bell inequalities.

\subsection{K- and B-Mesons Oscillations:}\label{sec:1.B.}
Mesons are the class of subatomic particles composed of quarks and anti-quarks bound states. These quark and anti-quark bound states are found as $\pi^+(u\bar{d})$, $\pi^-(u\bar{d})$, $K^0(\bar{s}d)$, $K^+(\bar{s}u)$, $\bar{K^0}(s\bar{d})$, $B^0_d(d\bar{b})$, $\bar{B^0_s}(\bar{s}b)$, etc. with charges 1, -1, 0, 1, 0, 0, 0. The neutral K meson, $K^0$, and B meson, $B^0$, and their respective antiparticles, $\bar{K^0}$ and $\bar{B^0}$ form a very exceptional quantum-mechanical two-state system, which has many key roles in elementary particle physics. Their physics has helped us to unveil frameworks to understand various aspects, such as the prediction of quark masses, the study of CKM matrices and their mixing ratios, and the amount of CP asymmetries in their respective decays. These mesons are generally produced in strong or electromagnetic interactions, which conserves their quark flavor, and they decay by weak interactions. Since the mesons are unstable, they eventually decay into lighter mesons having lighter quark masses and also into neutrinos, electrons, etc.; or in other words, they can decay semi-leptonically ($l$, $\nu_l$ and $\pi$) or hadronically ($\pi$).In semileptonic decays of K-mesons into the final state of pions with a particular distinct charge are $K^0\rightarrow l^+\nu_l\pi^-$ and $\hspace{0.15cm}\bar{K^0}\rightarrow l^-\bar{\nu_l}\pi^+$.
For hadronic decays, these K-mesons can result in two or three pions depending upon the state of superposition of $K^0$ and $\bar{K^0}$. Suppose we take the system of $K^0\bar{K^0}$ vacuum oscillation with the assumption of CP conservation in the decay of neutral kaons in the weak interaction. In this regard, the observed states of kaons will appear:
\begin{eqnarray}
 \resizebox{0.8\hsize}{!}{$|K_1^0\rangle=\frac{1}{\sqrt{2}}\left[|K^0\rangle+|\bar{K^0}\rangle \right],\hspace{0.125cm}
 |K_2^0\rangle=\frac{1}{\sqrt{2}}\left[|K^0\rangle-|\bar{K^0}\rangle \right]$}
\end{eqnarray}
$|K_1^0\rangle$ and $|K_2^0\rangle$ are the physical states that are available and are referred as short-lived mass eigenstates with proper lifetime $\tau_1$ and long-lived mass eigenstates with proper lifetime $\tau_2$ respectively. Each of them has a unique life-time. According to the assumptions these linear combination states will form the CP eigenstates.
\begin{eqnarray}
 \hat{C}\hat{P}|K_1^0\rangle=+|K_1^0\rangle\hspace{0.4cm}
 \hat{C}\hat{P}|K_2^0\rangle=-|K_2^0\rangle
\end{eqnarray}

Also, these states can be expressed back for $|K^0\rangle$ and $|\bar{K^0}\rangle$ states;
\begin{eqnarray}\label{eqn-18}
 \resizebox{0.8\hsize}{!}{$|K^0\rangle=\frac{1}{\sqrt{2}}\left[|K_1^0\rangle+|K_2^0\rangle \right],\hspace{0.25cm}
 |\bar{K^0}\rangle=\frac{1}{\sqrt{2}}\left[|K_1^0\rangle-|K_2^0\rangle \right]$}
\end{eqnarray}
Further if we wish to study the decay effects of kaons in a particular direction (say z) then by virtue of conservation of angular momentum the whole composite system of kaons will be correlated by the standard singlet configuration;
\begin{eqnarray}
|\Phi\rangle &=& \frac{1}{\sqrt{2}}\left[|K^0(z)\bar{K^0}(-z)\rangle+|\bar{K^0}(z)|K^0(-z)\rangle \right]\nonumber\\
             &=& \frac{1}{\sqrt{2}}\left[|K_2^0(z)K_1^0(-z)\rangle+|K_1^0(z)K_2^0(-z)\rangle \right]
\end{eqnarray}

The evolution of a $K^0\bar{K^0}$ oscillatory system in time can be studied by the time dependent Schrödinger equation:
\begin{eqnarray}
 i\frac{d}{dt}
 \begin{pmatrix}
 K^0(t)\\
 \bar{K^0}(t)
 \end{pmatrix}
 =
 \begin{pmatrix}
 H_{11} & H_{12}\\
 H_{21} & H_{22}
 \end{pmatrix}
 \begin{pmatrix}
 K^0(t)\\
 \bar{K^0}(t)
 \end{pmatrix}
\end{eqnarray}
The Hamiltonian matrix is composed of a hermitian $M$, the energy part of the oscillation and also an anti-hermitian $\Gamma$ which is responsible for the decay of oscillations and is equal to the inverse of proper lifetime;
\begin{eqnarray}
 H=M-\frac{i}{2}\Gamma
\end{eqnarray}
Therefore, the kaon state can be written as:
\begin{eqnarray}\label{eqn-19}
 \ket{\psi}=
 \begin{pmatrix}
 a\\
 b
 \end{pmatrix}
 =a
 \begin{pmatrix}
 1\\
 0
 \end{pmatrix}
 +b
 \begin{pmatrix}
 0\\
 1
 \end{pmatrix}
\end{eqnarray}
where, $\ket{K^0}=\begin{pmatrix}
1\\
0
\end{pmatrix}$
and $\ket{\bar{K^0}}=\begin{pmatrix}
0\\
1
\end{pmatrix}$.

\begin{eqnarray}
\text{ Thus,} \hspace{0.4cm}&\psi(t)=e^{-iHt}\psi(0)&\nonumber\\
\text{or} \hspace{0.3cm}
 &\begin{pmatrix}
 K^0(t)\\
 \bar{K^0}(t)
 \end{pmatrix}
 =e^{-\frac{\Gamma}{2}}e^{-iMt}
 \begin{pmatrix}
 K^0(0)\\
 \bar{K^0}(0)
 \end{pmatrix}&
\end{eqnarray}
If the system initially starts with only $\ket{K^0}$ state, then after time $t$, the propagation state will be:
\begin{eqnarray}
 \resizebox{0.87\hsize}{!}{$\ket{\psi(t)}=\frac{1}{\sqrt{2}}\left[ e^{-(im_1+\Gamma_1/2)}\ket{K^0_1}+e^{-(im_2+\Gamma_2/2)}\ket{K^0_2} \right]$}
\end{eqnarray}

Now, we can determine the corresponding probabilities;
\begin{eqnarray}\label{eqn-26}
 \resizebox{0.85\hsize}{!}{$P_{K^0}=\frac{1}{4}\left[ e^{-\Gamma_1t}+e^{-\Gamma_2t}+2e^{-(\Gamma_1+\Gamma_2)t/2}\cos{(\Delta m_K t)} \right]$}\\
  \resizebox{0.85\hsize}{!}{$P_{\bar{K^0}}=\frac{1}{4}\left[ e^{-\Gamma_1t}+e^{-\Gamma_2t}-2e^{-(\Gamma_1+\Gamma_2)t/2}\cos{(\Delta m_K t)} \right]$}\label{eqn-27}
\end{eqnarray}
where, $\Delta m_K=|m_1-m_2|$ is $K_1^0-K_2^0$ mass difference.
Equations (\ref{eqn-26}) and (\ref{eqn-27}); give the probabilities of state $\ket{K^0}$ remaining in the same itself and making the transition to its antiparticle state respectively \cite{Belusevic:1998pw}. In the same way, we can study the $B^0_d-\bar{B^0_d}$ and $B^0_s-\bar{B^0_s}$ oscillations (neutral $B$-mesons). We can do a similar time evolution and determine their probabilities with the corresponding neutral B-mesons parameters.

\subsection{Quantum Correlations in K- and B-Mesons oscillations:}
We have dealt with our systems so far considering a closed quantum system, however the systems that we encounter in real physical systems involve a significant interactions with its environment. So we implement the mathematical formalism of quantum operations which is an important tool for describing the dynamics of open quantum systems. We have to treat the states involved in our system as some mixed states instead of pure states and evolve them accordingly \cite{Banks:1983by}. The evolution of states should be carried through considering our system as a sub-system embedded in an exterior environment. Thus the evolution of the sub-system will differ from the system significantly \cite{nielsen_chuang_2010}. This will further lead to damping effects in our meson oscillation systems which originates in the form of decoherence in our open quantum states. This quantum decoherence effect can be understood using the Liouville-Lindblad formalism for open quantum systems \cite{Lindblad:1975ef}. The evolution of density matrix $\rho$ for the oscillatory system will be governed by;
\begin{eqnarray}
 \frac{\partial \rho}{\partial t}=-i\left[H, \rho\right]+D[\rho]
\end{eqnarray}
where $D[\rho]$ is the term that causes the non-unitary evolution of the system,and is termed as decoherence.  Here, our sub-system is $K^0-\bar{K^0}$ meson system, which is the principal system, and the environment it interacts with is attributed to our measurement action. The principal system and environment together forms a closed quantum system, and then our system-environment composite state will be characterized as $\rho \otimes \rho_{env}$. Now, performing a partial trace over the environment we will end up with the reduced state of the sub-system alone:
\begin{eqnarray}
 \mathcal{E}(\rho)=tr_{env}\left[U(\rho \otimes \rho_{env})U^{\dagger} \right] 
\end{eqnarray}
Here, $U$ is responsible for transforming an entire composite system. Suppose there is any other $\tilde{U}$ operator that only transforms the sub-system alone and does not interact with the environment. In that case, our final state will be as usual $\mathcal{E}(\rho)=\tilde{U}\rho\tilde{U}^{\dagger}$. This can be further put in an elegant representation known as operator-sum representation:
\begin{eqnarray}
\mathcal{E}(\rho) &=& \sum_i\bra{e_i}U\left[\rho\otimes\ket{e_0}\bra{e_0}\right]U^{\dagger}\ket{e_i}\nonumber\\
  &=& \sum_i \mathcal{K}_i\rho \mathcal{K}_i^{\dagger}
\end{eqnarray}

where $\mathcal{K}_i=\bra{e_i}U\ket{e_0}$ is an operator on the state space of the principal system. The operators \{$\mathcal{K}_i$\} are known as operation elements for the quantum operation $\mathcal{E}$. The \{$\ket{e_i}$\} is an orthonormal basis for the environment state space and $\rho_{env}=\ket{e_0}\bra{e_0}$ is its initial state. The operation elements satisfy the completeness property i.e., $tr(\mathcal{E}(\rho))=1$, this follows $\mathcal{K}_i$ operators to feature as trace preserve operations; $\sum_i \mathcal{K}_i^{\dagger}\mathcal{K}_i=I$.

These $\mathcal{K}_i$ are also known as Kraus operators \cite{1983LNP...190.....K}. Now, as our principal system is $K^0-\bar{K^0}$ oscillation system, we can choose our orthonormal basis as \{$\ket{K^0},\ket{\bar{K^0}},\ket{0}$\}, where $\ket{0}$ is a vacuum state. This basis will span the entire Hilbert space of our composite system \cite{Banerjee:2014vga,PhysRevA.72.032106, Naikoo:2019gme}.

\begin{eqnarray}\label{eqn-27}
 \ket{K^0}=
 \begin{pmatrix}
 1\\
 0\\
 0
 \end{pmatrix}
 ;
 \ket{\bar{K^0}}=
 \begin{pmatrix}
 0\\
 1\\
 0
 \end{pmatrix}
 ;
 \ket{0}=
 \begin{pmatrix}
 0\\
 0\\
 1
 \end{pmatrix}
\end{eqnarray}
Initially, when we treated the meson oscillations as a closed system, the Hilbert space $\mathcal{H}$ of our quantum system was spanned by the basis vectors; $\ket{K^0}$ and $\ket{\bar{K^0}}$ as given in equation (\ref{eqn-19}). Now, since we are considering an open quantum system for the meson oscillation system, the new Hilbert space $\mathcal{H}'$ will be a composite space comprising the original Hilbert space and the one corresponding to the environment $\mathcal{H}_{env}$; $\mathcal{H}'=\mathcal{H}\otimes \mathcal{H}_{env}$. This requires us to reconstruct the basis vectors in the composite Hilbert space. Thus, in this picture of an open quantum system, we can pick a vacuum state $\ket{0}$ along with the flavour states of mesons as basis vectors as shown in equation (\ref{eqn-27}). This further ensures that when we evolve the meson system in time, the flavour states mixes with a third state apart from mixing with each other; that accounts for the interaction with the
environment. If we proceed ahead with the general case of kaon oscillation by incorporating the CP violation measure \cite{Sarkar} in the oscillating states, then our new long and short-lived states will take the form:
\begin{eqnarray}
 \ket{K_S}=\frac{1}{\sqrt{1+|\epsilon|^2}}\left( \ket{K_1^0}+\epsilon\ket{\bar{K_2^0}} \right)\nonumber\\
  \ket{K_L}=\frac{1}{\sqrt{1+|\epsilon|^2}}\left(\epsilon \ket{K_1^0}+\ket{\bar{K_2^0}} \right)
\end{eqnarray}

where $\epsilon$ is the CP violating parameter.
Then the evolution of density matrix for our principal system is:
\begin{eqnarray}
 \rho(t)=\sum_{i=0}\mathcal{K}_i(t)\rho(0)\mathcal{K}_i^{\dagger}(t)
\end{eqnarray}
Kraus operators here are expressed in the following way \cite{Naikoo:2018vug}:

\begin{eqnarray}\label{eqn-35}
 \mathcal{K}_0 &=& \ket{0}\bra{0}\nonumber\\
 \mathcal{K}_1 &=& \mathcal{C}_{1+}(\ket{K^0}\bra{K^0}+\ket{\bar{K^0}}\bra{\bar{K^0}})\nonumber\\
 &&
 +\mathcal{C}_{1-}(\frac{1+\epsilon}{1-\epsilon}\ket{K^0}\bra{\bar{K^0}}+\frac{1-\epsilon}{1+\epsilon}\ket{\bar{K^0}}\bra{K^0})\nonumber\\
 \mathcal{K}_2 &=& \mathcal{C}_{2}(\frac{1}{1+\epsilon}\ket{0}\bra{K^0}+\frac{1}{1-\epsilon}\ket{0}\bra{\bar{K^0}})\nonumber\\
 \mathcal{K}_3 &=& \mathcal{C}_{3+}\frac{1}{1+\epsilon}\ket{0}\bra{K^0}+\mathcal{C}_{3-}\frac{1}{1-\epsilon}\ket{0}\bra{\bar{K^0}}\nonumber\\
  \mathcal{K}_4 &=& \mathcal{C}_{4}(\ket{K^0}\bra{K^0}+\ket{\bar{K^0}}\bra{\bar{K^0}}\nonumber\\
  &&
  +\frac{1+\epsilon}{1-\epsilon}\ket{K^0}\bra{\bar{K^0}}+\frac{1-\epsilon}{1+\epsilon}\ket{\bar{K^0}}\bra{K^0})\nonumber\\
   \mathcal{K}_5 &=& \mathcal{C}_{5}(\ket{K^0}\bra{K^0}+\ket{\bar{K^0}}\bra{\bar{K^0}}\nonumber\\
 &&
  -\frac{1+\epsilon}{1-\epsilon}\ket{K^0}\bra{\bar{K^0}}-\frac{1-\epsilon}{1+\epsilon}\ket{\bar{K^0}}\bra{K^0})
 \end{eqnarray}
 
 The coefficients are as follows:
\begin{eqnarray}\label{eqn-36}
 \mathcal{C}_{1\pm} &=& \frac{1}{2}\left[e^{-(2im_S+\Gamma_S+\lambda)t/2}\pm e^{-(2im_L+\Gamma_L+\lambda)t/2}\right] \nonumber\\
 \mathcal{C}_{2} &=& \sqrt{\frac{1+|\epsilon|^2}{2}\left(1-e^{-\Gamma_St}-\delta_L^2\frac{|1-e^{-(\Gamma+\lambda-i\Delta m)t}|^2}{1-e^{-\Gamma_Lt}}\right)}\nonumber\\
 \mathcal{C}_{3\pm} &=& \sqrt{\frac{1+|\epsilon|^2}{2(1-e^{-\Gamma_Lt})}}
 \left[1-e^{-\Gamma_Lt}\pm (1-e^{-(\Gamma+\lambda-i\Delta m)t})\right] \nonumber\\
 \mathcal{C}_{4} &=& \frac{e^{-\Gamma_St/2}}{2}\sqrt{1-e^{-\lambda t}}\nonumber\\
 \mathcal{C}_{5} &=& \frac{e^{-\Gamma_Lt/2}}{2}\sqrt{1-e^{-\lambda t}}
\end{eqnarray}

Now, for the evolution of density matrices for the oscillations of $K^0-\bar{K^0}$, we can evolve it by the help of these Kraus operators. Here, $\lambda$ denotes the decoherence parameter that accounts for the decoherence effect in the process of evolution of the meson state by Kraus operators. The decoherence effect is a phenomenon that makes the system mixed with the external environment. The system lacks coherence and loses information to the environment as it evolves through time \cite{Schlosshauer:2019ewh}.\\
When the system was initially at $\ket{K^0}$ state:
\begin{eqnarray}
 \rho_{K^0}(t) &=& \sum_i\mathcal{K}_i \rho_{K^0}(0)\mathcal{K}_i^{\dagger}\nonumber\\
                &=& \mathcal{K}_0 \rho_{K^0}(0)\mathcal{K}_0^{\dagger}+\mathcal{K}_1 \rho_{K^0}(0)\mathcal{K}_1^{\dagger}+...+\mathcal{K}_5 \rho_{K^0}(0)\mathcal{K}_5^{\dagger}\nonumber\\
\end{eqnarray}

Similarly, if the system was initially at $\ket{\bar{K^0}}$ state:
\begin{eqnarray}
 \rho_{\bar{K^0}}(t) &=& \sum_i\mathcal{K}_i\rho_{\bar{K^0}}(0)\mathcal{K}_i^{\dagger}
\end{eqnarray}
The  $\rho_{K^0}(0)$  and $\rho_{\bar{K^0}}(0)$ are the corresponding initial density operators for the pure states $\ket{K^0}\bra{K^0}$ and $\ket{\bar{K^0}}\bra{\bar{K^0}}$ respectively. Using these, we can construct the respective evolved density operators (as solved in Appendix [\ref{append:1}]) from equations (\ref{eqn-35}) and (\ref{eqn-36}) in the following form:

\begin{eqnarray}\label{eqn-39}
 \resizebox{1.0\hsize}{!}{$\rho_{K^0}(t)=
 \frac{1}{2}e^{-\Gamma t}
 \begin{pmatrix}
 a_{ch}+e^{-\lambda t}a_c & (\frac{1-\epsilon}{1+\epsilon})^*(-a_{sh}-ie^{-\lambda t}a_s)  & 0 \\
( \frac{1-\epsilon}{1+\epsilon})(-a_{sh}+ie^{-\lambda t}a_s) & |\frac{1-\epsilon}{1+\epsilon}|^2(a_{ch}-e^{-\lambda t}a_c) & 0  \\
 0 & 0 & \rho_{33}(t) \nonumber \\
 \end{pmatrix}$}
\end{eqnarray}

\begin{eqnarray}
 \resizebox{1.0\hsize}{!}{$\rho_{\bar{K^0}}(t)=
 \frac{1}{2}e^{-\Gamma t}
 \begin{pmatrix}
|\frac{1+\epsilon}{1-\epsilon}|^2 (a_{ch}+e^{-\lambda t}a_c) & (\frac{1+\epsilon}{1-\epsilon})(-a_{sh}+ie^{-\lambda t}a_s)  & 0 \\
(\frac{1+\epsilon}{1-\epsilon})^*(-a_{sh}-ie^{-\lambda t}a_s) & (a_{ch}-e^{-\lambda t}a_c) & 0  \\
 0 & 0 & \tilde{\rho_{33}}(t) \\
 \end{pmatrix}$}\nonumber\\
\end{eqnarray}

The notations used for the matrix elements denotes these following terms:
\begin{itemize}
    \item $a_{ch}=\cosh{\left(\frac{\Delta\Gamma t}{2}\right)}$ 
    \item $a_{sh}=\sinh{\left(\frac{\Delta\Gamma t}{2}\right)}$
    \item $a_c=\cos{(\Delta mt)}$
    \item $a_s=\sin{(\Delta mt)}$
    \item $\epsilon=\text{CP violating parameter}$
    \item $\Delta \Gamma=\Gamma_S-\Gamma_L$, difference of decay widths
    \item $\Gamma=\frac{1}{2}(\Gamma_L+\Gamma_S)$ is the mean decay width
    \item $\Delta m=m_L-m_S$, mass difference
\end{itemize}

Equations (\ref{eqn-39}), give density matrices at time $t$, whose diagonal elements give probabilities for making transition from one state to the other at time $t$, and all other off-diagonal elements describe the superposition states of all states involved. If the initial system starts out with state $\ket{\bar{K^0}}$ and probability of remaining in the same state at time $t$ will therefore be given by;

\begin{eqnarray}
P_{\Bar{K}^0\Bar{K}^0}=\frac{e^{-\Gamma t}}{2}\left[\cosh{\left(\frac{\Delta\Gamma t}{2}\right)}+e^{-\lambda t}\cos{(\Delta mt)} \right]
\end{eqnarray}

Similarly, other probabilities at time $t$ can be found from the density matrices in equations (\ref{eqn-39}).

\section{Analysis for $3$-measurement LGI parameter $K_3$}
The two-time correlations which is defined in equation (\ref{eqn-4}), can now be redefined in the context of Kraus operators \cite{PhysRevA.88.022104,PhysRevLett.71.3235, PhysRevA.54.1798, PhysRevLett.101.090403}. 
\begin{eqnarray}\label{eqn-41}
   C(t_i,t_j) &=& p(^+t_i)q(^+t_j|^+t_i)-p(^+t_i)q(^-t_j|^+t_i)\nonumber\\
   &&
   -p(^-t_i)q(^+t_j|^-t_i)+p(^-t_i)q(^-t_j|^-t_i)\nonumber\\
\end{eqnarray}

Here, $p(^at_i)$ gives the probability of getting the result $a=\pm1$
at $t_i$ means initially the system has this probability of remaining in one of the states corresponding to the value of $a=\pm1$ and $q(^bt_j|^at_i)$ gives the conditional probability (or also known as transition probability) which describes that the system has this probability of making transition to another state or remaining in that state (i.e., $b=\pm1$) at $t_j$ given that the system in the either of one state corresponding to $a=\pm 1$ at $t_i$. We know that the probability of a quantum state ($\ket{\psi}=\sum_i c_i\ket{\phi_i}$) to be in one of its basis states is equal to the expectation value of projector operator corresponding to that basis state, i.e., $|c_i|^2=\langle \hat{P_i} \rangle$, where $\hat{P_i}$ is the projector operator for that basis state. This expectation value further, with its density operator, can be expressed as $\langle \hat{P_i} \rangle= Tr\{\hat{P_i}\hat{\rho}\}$. Thus, we can define the probabilities in equation (\ref{eqn-41}) as:
\begin{eqnarray}
   p(^at_i)=Tr\{\Pi^a\rho(t_i) \}=Tr\{\Pi^a\sum_{\mu}\mathcal{K}_{\mu}(t_i)\rho(0)\mathcal{K}_{\mu}^{\dagger}(t_i) \}\nonumber\\
\end{eqnarray}
where, $\Pi^a$ is the density operator for the dichotomic operator $\hat{Q}$ depending on the value of $a=\pm 1$. Here, we can take $a=+1$ for $\ket{K^0}$ and $a=-1$ for $\ket{\bar{K^0}}$, $\ket{0}$ states for $Q$ observable and LGI parameter $K_3$ in equation(\ref{eqn-8}) will still remain the same \cite{PhysRevLett.113.050401}. Further, dichotomic operator can be defined as $\Pi=\Pi^+-\Pi^-=\Pi_{K^0}-(\Pi_{\bar{K^0}}+\Pi_0)$, where $\Pi_x=\ket{x}\bra{x}$ \cite{Wilde}. Also, since $|c_i|^2=\langle \hat{P_i} \rangle$, i.e., $\bra{\psi_i}\rho\ket{\psi_i}=Tr\{\hat{P_i}\hat{\rho} \}$ or in our case $\bra{\psi_a}\rho\ket{\psi_a}=Tr\{\Pi^a\rho\}$, then we can show;
\begin{eqnarray}
   \rho^a &=& \ket{\psi_a}\bra{\psi_a}=\ket{\psi_a}\left(\frac{\bra{\psi_a}\rho\ket{\psi_a}}{Tr\{\Pi^a\rho\}} \right)\bra{\psi_a}\nonumber\\
   \implies \rho^a(t_i) &=& \frac{\Pi^a\rho(t_i)\Pi^a}{Tr\{\Pi^a\rho(t_i)\}}\nonumber\\
   &=& \frac{\Pi^a\sum_{\mu}\mathcal{K}_{\mu}(t_i)\rho(0)\mathcal{K}_{\mu}^{\dagger}(t_i)\Pi^a}{p(^at_i)}
\end{eqnarray}

In order to compute the probability $q(^bt_j|^at_i)$, we have to further evolve this density matrix to $t_j$, which would go like $\sum_{\nu}\mathcal{K}_{\nu}(t_j-t_i)\rho^a(t_i)\mathcal{K}_{\nu}^{\dagger}(t_j-t_i)$, hence, now in general probability is described as:
\begin{eqnarray}
&&
   p(^at_i)q(^bt_j|^at_i)=\nonumber\\
   &&
  \resizebox{.98\hsize}{!}{$Tr\left\{\Pi^b\sum_{\nu,\mu}\mathcal{K}_{\nu}(t_j-t_i)\Pi^a\mathcal{K}_{\mu}(t_i)\rho(0)\mathcal{K}_{\mu}^{\dagger}(t_i)\Pi^a\mathcal{K}_{\nu}^{\dagger}(t_j-t_i)
\right\}$}\nonumber\\
\end{eqnarray}

Finally, using this expression of probability we can compute all terms in equation (\ref{eqn-41}) which will yield the two-time correlation function $C(t_i,t_j)$ in the following form (Appendix [\ref{append:2}]):
\begin{eqnarray}\label{eqn-45}
   C(t_i,t_j)=1-2p(^+t_i)-2p(^+t_j)+4Re[g(t_i,t_j)]
\end{eqnarray}
where,
\begin{eqnarray}
   g(t_i,t_j)=Tr\left\{\Pi^+\sum_{\nu}\mathcal{K}_{\nu}(t_j-t_i)\Pi^+\rho(t_i)\mathcal{K}_{\nu}^{\dagger}(t_j-t_i) \right\}\nonumber\\
\end{eqnarray}
To simplify our work, we can now invoke stationary state assumption due to which two-time correlations will depend only on the time difference between two instants and this is also what noninvasive measurement of LGI formalism claims \cite{PhysRevA.54.1798}. Also, the probabilities $p(^at_i)$ will remain independent of time since the density operators won't change over time. Also, this will lead to $g(t_i,t_j)$ expression to depend only on the time interval between the initial and final instant of time, i.e.,  as $g(t_j-t_i)$. So, now the correlation function can be given as:
\begin{eqnarray}
  C(t_i,t_j)=1-4p^++4Re[g(t_j-t_i)]
\end{eqnarray}

If we lay down equal time interval settings as $t_1=0$, $t_2=\tau$, $t_3=2\tau$ and if our system starts out with initial state $\ket{\bar{K^0}}$ then $p(^at_i)=P_{\Bar{K}^0K^0}$, (from equations (\ref{eqn-39}) we have);
\begin{eqnarray}
 P_{\Bar{K}^0K^0}=\frac{e^{-\Gamma t}}{2}\left|\frac{p}{q}\right|^2\left[\cosh{\left(\frac{\Delta\Gamma t}{2}\right)}+e^{-\lambda t}\cos{(\Delta mt)} \right]
\end{eqnarray}

where $\frac{p}{q}=\frac{1+\epsilon}{1-\epsilon}$.
The correlation function $C_{12}$ in all these settings will then become:
\begin{eqnarray}
 C_{12}=1-4P_{\bar{K^0}K^0}+4Re[g(\Delta t)]
\end{eqnarray}
where,
\begin{eqnarray}
 g(t_1,t_2)=2P_{\bar{K^0}K^0}P_{\bar{K^0}\bar{K^0}}+\left|\frac{p}{q}\right|^2\frac{e^{-2\Gamma\Delta t}(e^{-2\lambda\Delta t}-1)}{4}\nonumber\\
\end{eqnarray}

The final form of LGI function under all these settings can be written as following \cite{Naikoo:2018vug, Naikoo:2019eec, Naikoo:2017fos}:

\begin{eqnarray}
K_3 = 1-4P_{\Bar{K}^0K^0}(\Delta t)+8P_{\Bar{K}^0K^0}(\Delta t)P_{\Bar{K}^0\Bar{K}^0}(\Delta t)\nonumber\\
+\left| {{\frac{p}{q}}}\right|^2e^{-2\Gamma\Delta t}\left( e^{-2\lambda\Delta t}-1\right)
\end{eqnarray}

\begin{figure}[H]
\centering
\includegraphics[height=6.5cm, width=7cm]{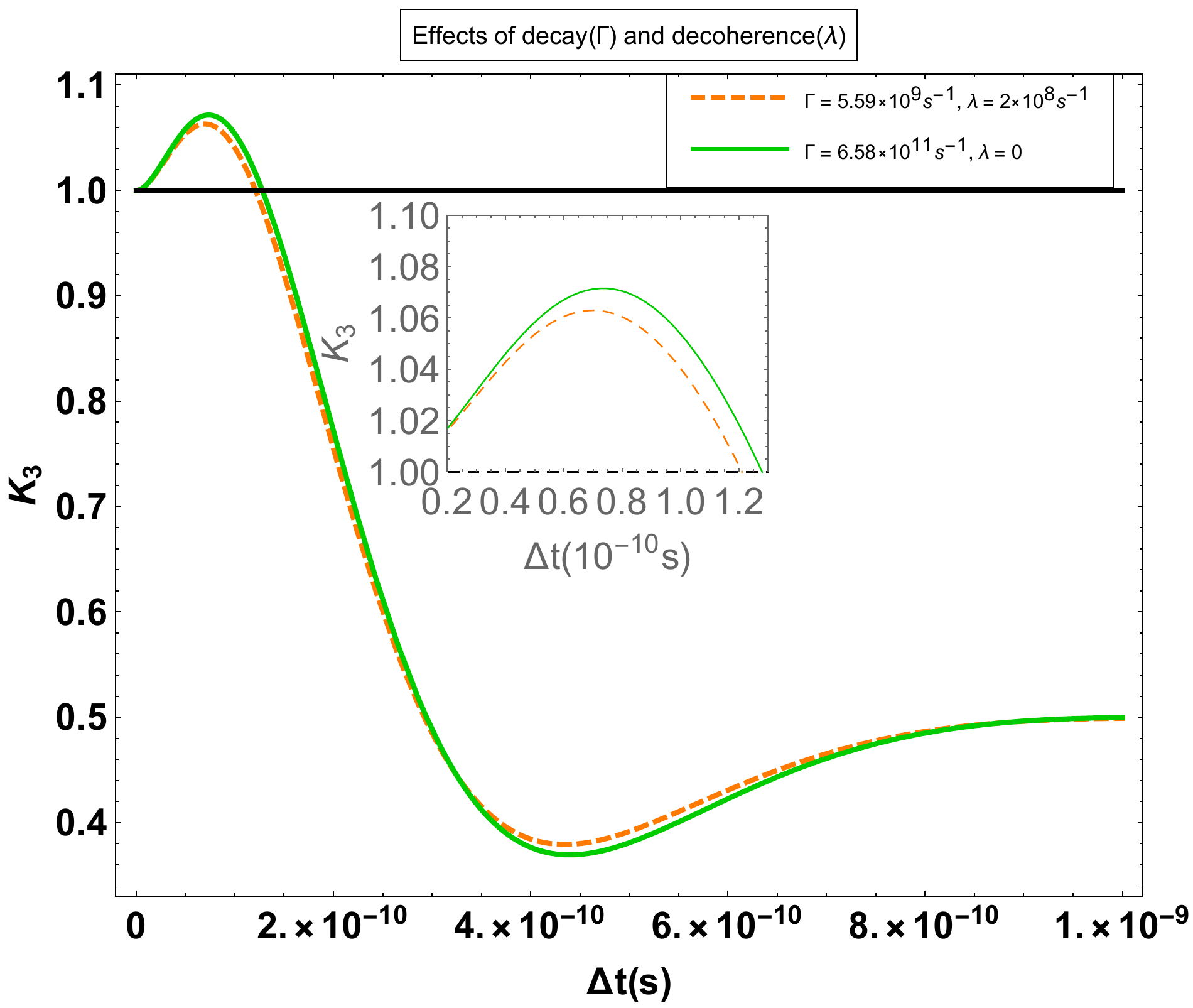}
\caption{Plot of $K_3$ vs $\Delta t$ when the system initially started out with $\ket{\bar{K^0}}$ state. The red shaded curve shows the variation for oscillations with decay and decoherence (i.e, $\Gamma\neq0$;$\lambda\neq0$). The green bold curve shows the variation for oscillations with decay and without decoherence (i.e, $\Gamma\neq0$;$\lambda=0$). The horizontal black line denotes the maximum value of $K_3$ allowed by LGI. The CP violating parameter is taken to be $\frac{q}{p}=\frac{1-\epsilon}{1+\epsilon}$  where $|\epsilon|=2.28\times10^{-3}$ \cite{HFLAV:2016hnz}, $\Delta \Gamma=1.1174\times10^{10}s^{-1}$ and the value of meson mixing parameter $\Delta m=5.302\times10^{9}s^{-1}$ ~\cite{HFLAV:2016hnz}. In the middle, there is a zoomed view of the peak region of plot where horizontally it varies from $0.25\times10^{-10}s$ to $1.25\times10^{-10}s$ and vertically from 1.00 to 1.10, which gives further validation of how distinct is behavior of both the cases at a closer level.}
\label{Fig.1-}
\end{figure}

We see from FIG. \ref{Fig.1-}, that there is an apparent violation of the upper bound limit of $K_3$ which goes beyond one specifically for almost all the values between $\Delta t=0$ and $2\times10^{-10}s$ depicted on the plot. The decoherence parameter $\lambda$, which defines the strength of the interaction between the system and its environment is fixed to $2\times 10^8/s$~\cite{HFLAV:2016hnz,Patrignani_2016,DAmbrosio:2006hes}. The behavior of the green curve represents only the effect of decay, and the orange curve represents the combined effect of decay and decoherence when the value of $\Delta t$ is around or equal to $10^{-10}s$, a maximum violation is attained for both cases. The zoom-in view shows the violation is more significant when no decoherence effect is involved. Thus, the system of meson oscillation tends to exhibit maximal temporal correlation when its flavor eigenstates are coherent.

Suppose we turn off both the decay and decoherence effects governing these oscillations, leading to a different scenario. The expressions for probabilities and LGI function will change as:
\begin{itemize}
    \item Removing decay effects:\\
    \begin{eqnarray}
    P_{\Bar{K}^0\Bar{K}^0}=\frac{1}{2}\left[1+e^{-\lambda t}\cos{(\Delta mt)} \right]
    \end{eqnarray}
    \begin{eqnarray}
    K_3 = 1-4P_{\Bar{K}^0K^0}(\Delta t)+8P_{\Bar{K}^0K^0}(\Delta t)P_{\Bar{K}^0\Bar{K}^0}(\Delta t)\nonumber\\
+\left| {{\frac{p}{q}}}\right|^2\left( e^{-2\lambda\Delta t}-1\right)
    \end{eqnarray}

\end{itemize}

\begin{figure}[H]
    \centering
    \includegraphics[height=6.5cm, width=7cm]{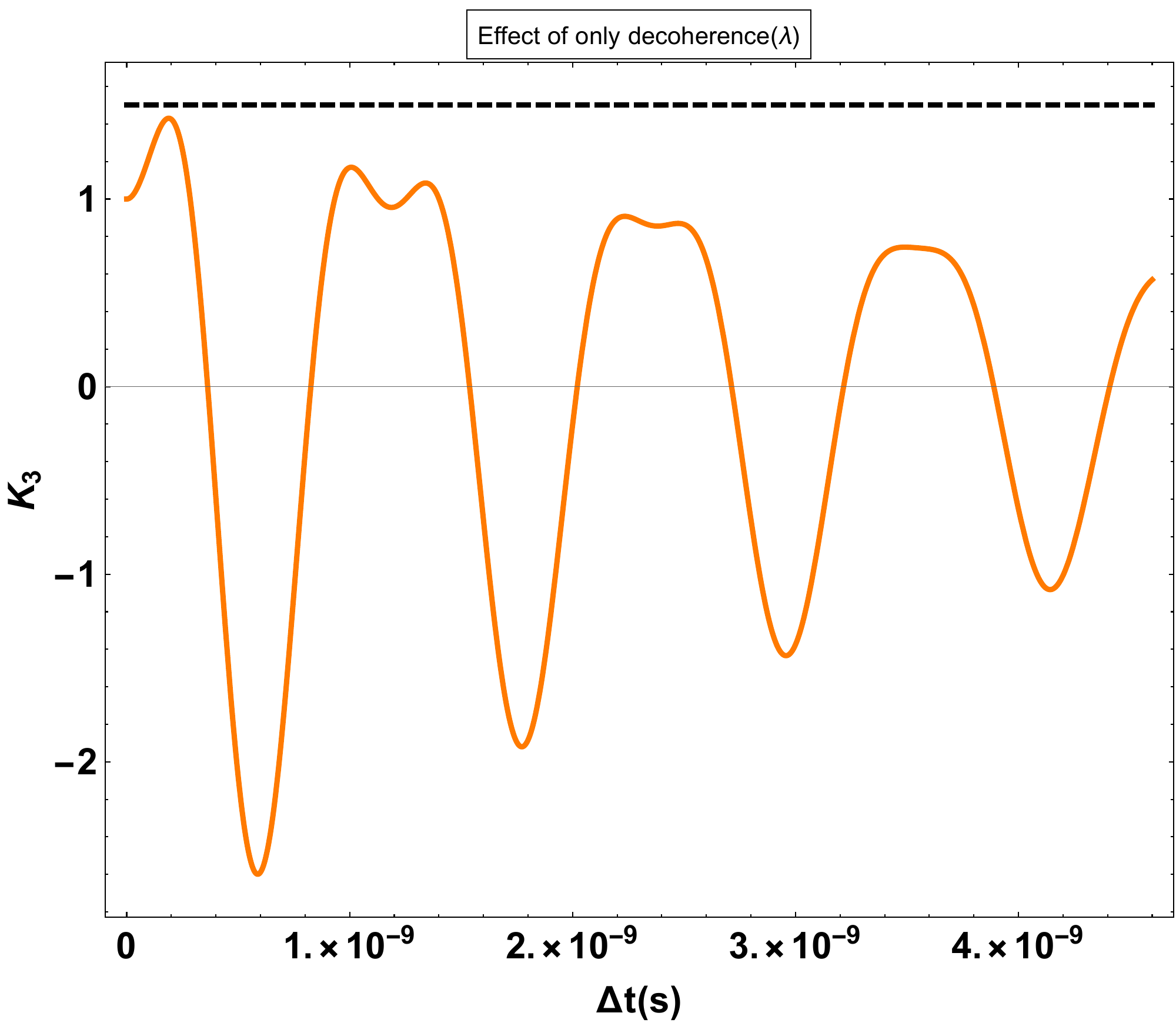}
    \caption{Plot for $K_3$ vs $\Delta t$ when decay effect is completely neglected and decoherence is taken into account. The horizontal shaded line denotes the maximum quantum mechanical violation for $K_3$.}
    \label{Fig.2-}
\end{figure}

\begin{itemize}
    \item Removing decay and decoherence effects:\\
    \begin{eqnarray}
    P_{\Bar{K}^0\Bar{K}^0}=\frac{1}{2}\left[1+\cos{(\Delta mt)} \right]
    \end{eqnarray}
    \begin{eqnarray}
     K_3 = 1-4P_{\Bar{K}^0K^0}(\Delta t)+8P_{\Bar{K}^0K^0}(\Delta t)P_{\Bar{K}^0\Bar{K}^0}(\Delta t)
    \end{eqnarray}
\end{itemize}

\begin{figure}[H]
    \centering
    \includegraphics[height=6.5cm, width=7cm]{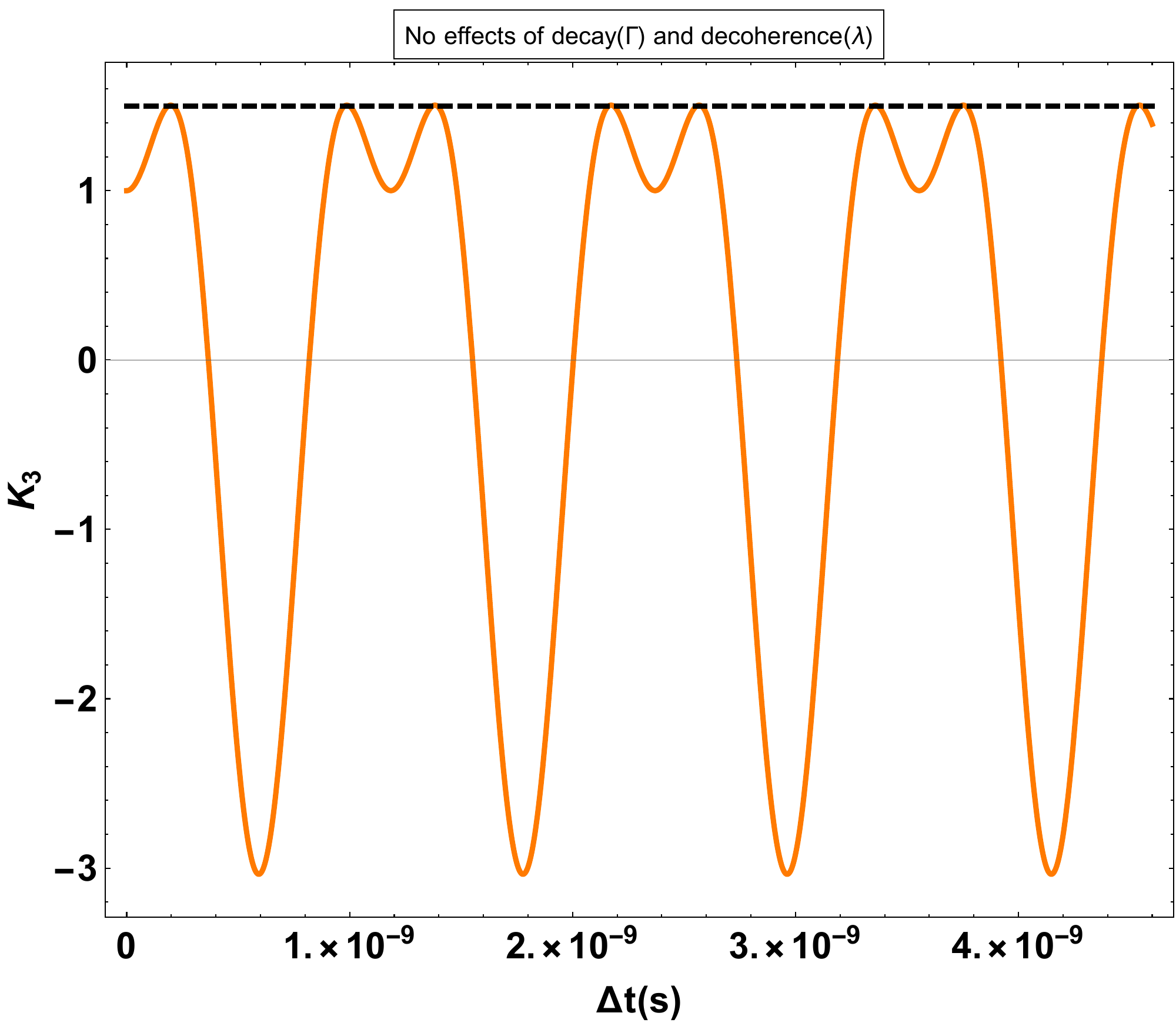}
    \caption{Plot for $K_3$ vs $\Delta t$ when both decay and decoherence effects are completely neglected.The horizontal shaded line denotes the maximum quantum mechanical violation for $K_3$.}
    \label{Fig.3-}
\end{figure}

From FIG. \ref{Fig.2-} \& FIG. \ref{Fig.3-}, intuitively, we get the violations again. Here, we can see the oscillatory behavior of probabilities is still preserved in both cases. This feature exists because the decay effect is completely neglected. However, in the latter case, the decoherence effect is neglected, making it look like an ideal oscillatory system. Thus, again, the meson system is completely coherent, and we see the maximal violation at successive intervals of $\Delta t$ where it attains the upper saturation bound of $1.5$. In the other case, we see the behavior of the curve is attenuating where the strength of amplitude decreases successively with the same $\Delta t$ interval. This happens because the temporal frequency or mixing parameter of mesons $\Delta m$ is kept fixed. The upper limit maximum violation in the FIG. \ref{Fig.2-} is achieved around when the value of $\Delta t=0.2\times10^{-9}s$ which is just below $K_3=1.5$ and the violation gradually decreases at successive intervals of $\Delta t$. In contrast, FIG. \ref{Fig.3-} depicts the maximum violation always repeate after every successive $\Delta t$. If this particular sort of behavior persists in mesons, we can exploit it in various interferometry examinations leading to the test of noninvasive measurability \cite{PRXQuantum.3.010307}. We can properly make an illustrative comparison in the following figure, summarizing all the different cases in a single picture.

Also, when decay and decoherence effect is studied individually, as in FIG. \ref{Fig.1-} (bold green curve) and in FIG. \ref{Fig.2-} shows two different responses of $K_3$.  The violation of inequality for the former case occurred at the interval of the order $(10^{-10}s)$, but for the latter, it is found at the interval of the order $(10^{-9}s)$. This distinctly clarifies that decay and decoherence times are different and have different physical origins. With that, when only the decay effect is present in the system, the maximum violation is attained below the value of $K_3=1.1$. For only the decoherence effect, the maximum violation is attained below the value of quantum bound $K_3=1.5$.

\begin{figure}[H]
    \centering
    \includegraphics[height=7cm, width=9cm]{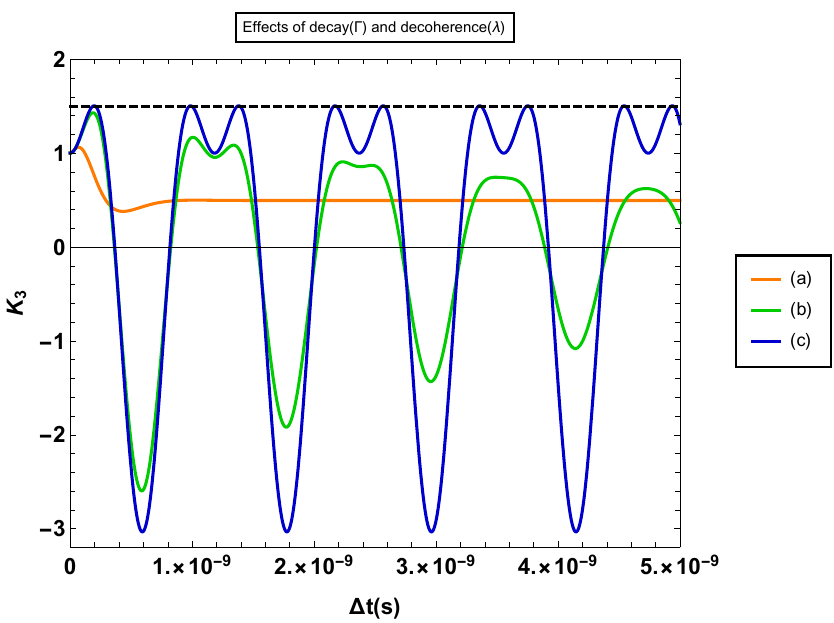}
    \caption{Plot for $K_3$ vs $\Delta t$ for three different cases, i.e., (a). oscillations with decay and decoherence given by red ($\Gamma\ne0$ \& $\lambda\ne0$), (b). oscillations without decay and with decoherence given by green ($\Gamma=0$ \& $\lambda\ne0$) and (c). oscillations without decay and decoherence given by blue ($\Gamma=0$ \& $\lambda=0$). The horizontal shaded line denotes the maximum quantum mechanical violation for $K_3$.}
    \label{Fig.4-}
\end{figure}

Now, FIG. \ref{Fig.4-} indicates that the practical situation of oscillation, i.e., when both decay and decoherence are involved, differs significantly from the response of other scenarios. We saw it in FIG. \ref{Fig.1-} maximum violation is achieved only for specific interval $\Delta t$; otherwise, it always saturates to a lower limit of LGI $K_3=0.5$ for intervals $\Delta t\ge 10^{-9}s$. Also, from FIG \ref{Fig.1-}, we can verify that the upper quantum bound, $K_3=1.5$, was never attained in this case, far from the upper quantum saturation mark. The important outtake which we can derive from FIG. \ref{Fig.1-} \& \ref{Fig.4-} is whenever we have a system that is entirely devoid of decay effects $\Gamma=0$ (or practically, we can say observing a system at a very negligible time interval which is just comparable to their lifetime) then the system consistently attain the maximal violation of $K_3=1.5$ or reaches the maximum value just below it. Also, when we completely cut down the interaction of the environment ($\lambda=0$), it acts as a coherent oscillation system. This allows one to infer that if there is no quantum decoherence in flavor oscillations of the meson system, then it will always saturate the temporal \textit{Tsirelson bound} \cite{Wang2017BeyondTT} for at least a specific temporally separated $\Delta t$ and correlated flavor states. This means that these temporally correlated states will exhibit maximal entanglement in time compared to any other cases. This is just like the maximal violation of Bell inequality for a given bipartite two-qubit state \cite{XIANG}.\\

So far, we have observed all about the $K$-mesons; now, we can turn to $B_d$-mesons. Here, too, we will have the same behavior for variation of $K_3$. But still, there can be some subtleties to watch out for since both mesons have different quark contents, thereby making them have different decay modes, mixing parameters, etc.  

\begin{figure}[H]
\centering
\includegraphics[height=7cm, width=9cm]{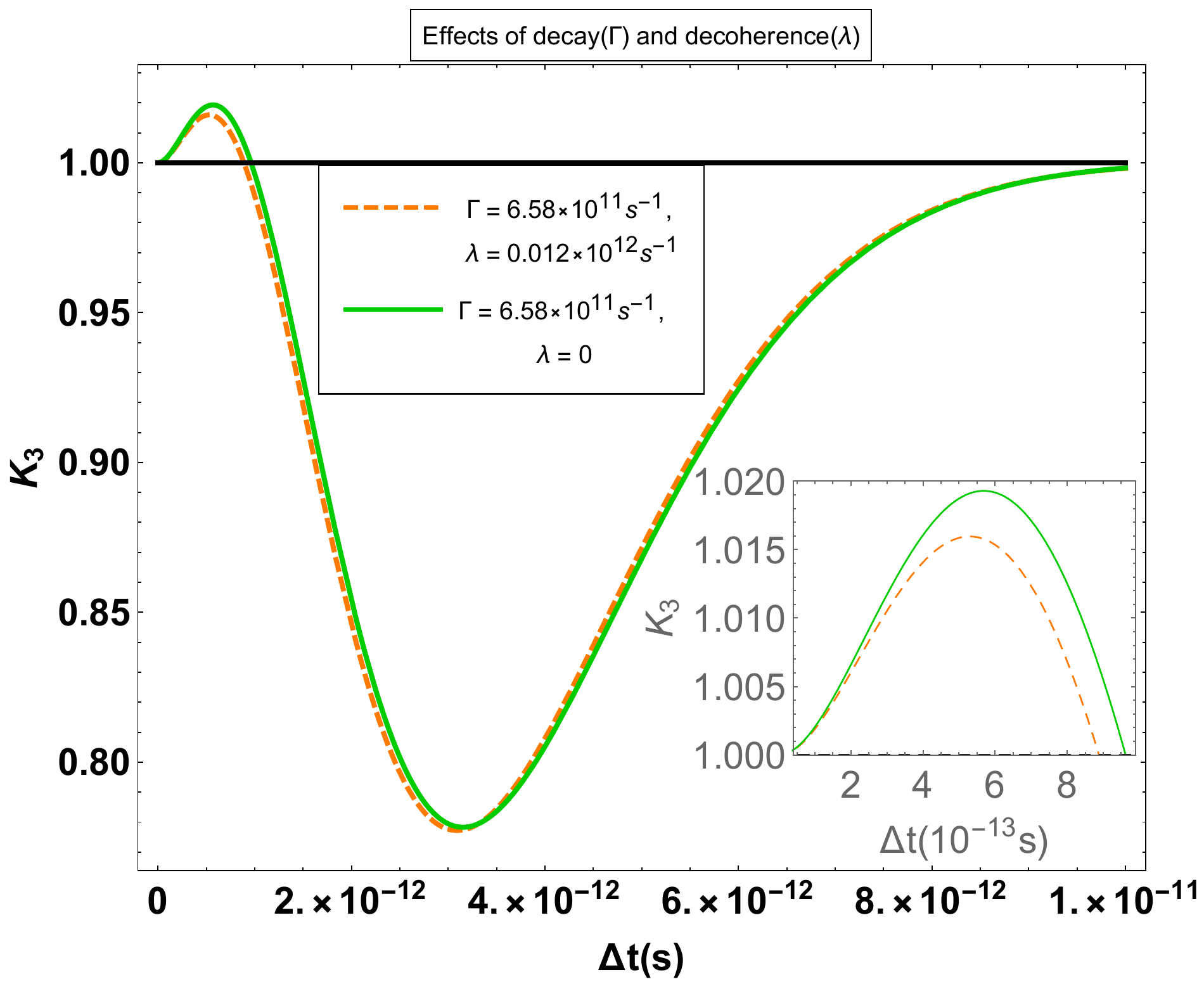}
    \caption{Plot for $K_3$ vs $\Delta t$. Again, the red shaded curve and green bold curve corresponds to $\Gamma\neq0$; $\lambda\neq0$ and $\Gamma\neq0$; $\lambda=0$ respectively. The horizontal black line denotes the maximum value of $K_3$ allowed by LGI. The CP violating parameter is taken to be $\left|\frac{q}{p}\right|=1.010$ \cite{DAmbrosio:2006hes}. Other values are taken as $\Delta \Gamma=0$ and $\Delta m=0.5064\times10^{12}s^{-1}$ \cite{Olive_2016}}
    \label{Fig.5-}
\end{figure}

\begin{figure}[H]
    \centering
    \includegraphics[height=7cm, width=9cm]{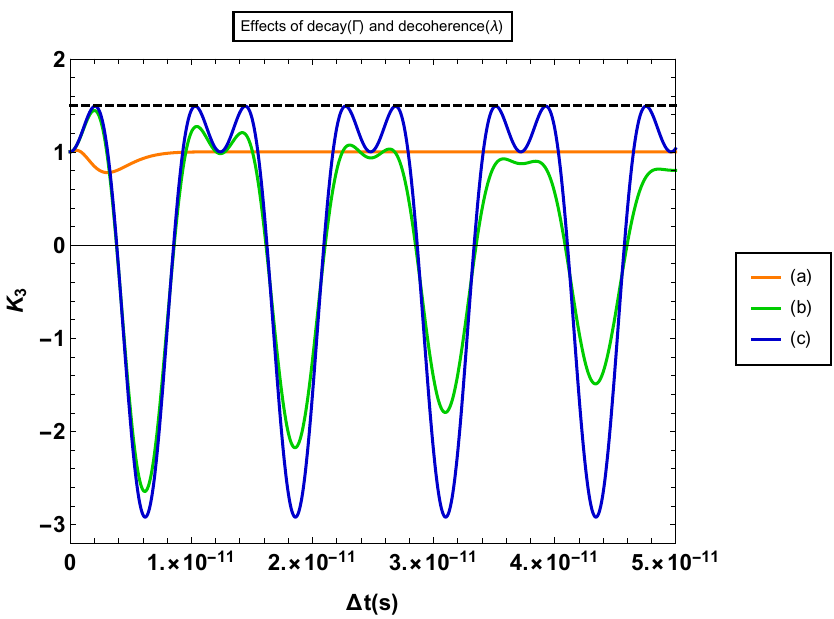}
    \caption{Plot for $K_3$ vs $\Delta t$ for three different cases: (a). $\Gamma\ne0$ \& $\lambda\ne0$, (b). $\Gamma=0$ \& $\lambda\ne0$ and (c). $\Gamma=0$ \& $\lambda=0$. The horizontal shaded line denotes the maximum quantum mechanical violation for $K_3$.}
    \label{Fig.6-}
\end{figure}

Now, FIG. \ref{Fig.5-} and \ref{Fig.6-} responds similarly to what we found in $K$-mesons. Here also, we see whenever there is complete coherence in the oscillations or practically if we study the system for minimal time intervals compared to the lifetimes, then we find the system maximally violates the LGI or saturates the \textit{Tsirelson bound} (upper quantum bound; $K_3=1.5$). Again, here, the individual study of decay and decoherence has shown two distinct variations of $K_3$. The maximum violation obtained around the order of $ 10 {-12}s$ and $ 10 {-11}s$ for only decay and only decoherence effect, respectively, in $B_d$-meson oscillatory system. This again validates that decay and decoherence time stems from different physical origins. However, in $B_d$-mesons, when both decay and decoherence are involved, there is little different behavior, unlike in $K$-mesons. The upper limit of $K_3$ value as predicted by LGI (or also $K_3>1$) for K-mesons is only reached for very short instants, and otherwise, it takes the constant value of $K_3=0.5$ for all the higher intervals $\Delta t$. But in $B_d$-mesons, except for some small interval, its value always stays at 1.

If we digress back to the study of meson oscillation in the context of CP conservation in the section [\ref{sec:1.B.}], we can refer to their probabilities and construct the correlation and LGI function accordingly. Again invoking the equal time interval measurement assumption on the $K$-meson oscillation when initially it was started with $\ket{K^0}$ state, then we would yield the following results from the definitions as stated in equations (\ref{eqn-4}) and (\ref{eqn-6});

\begin{eqnarray}
 C_{12}(\Delta t)= 2(P_{K^0K^0}(\Delta t)-P_{K^0\bar{K^0}}(\Delta t))
  \end{eqnarray}
  
 \begin{eqnarray}
 \text{ \& }
 K_3=2 C_{12}(\Delta t)- C_{12}(2\Delta t)
\end{eqnarray}
where $P_{K^0K^0}$ and $P_{K^0\bar{K^0}}$ are the expressions of probabilities as given in equations (\ref{eqn-26}) and (\ref{eqn-27}).

\begin{figure}[H]
    \centering
    \includegraphics[height=7cm, width=9cm]{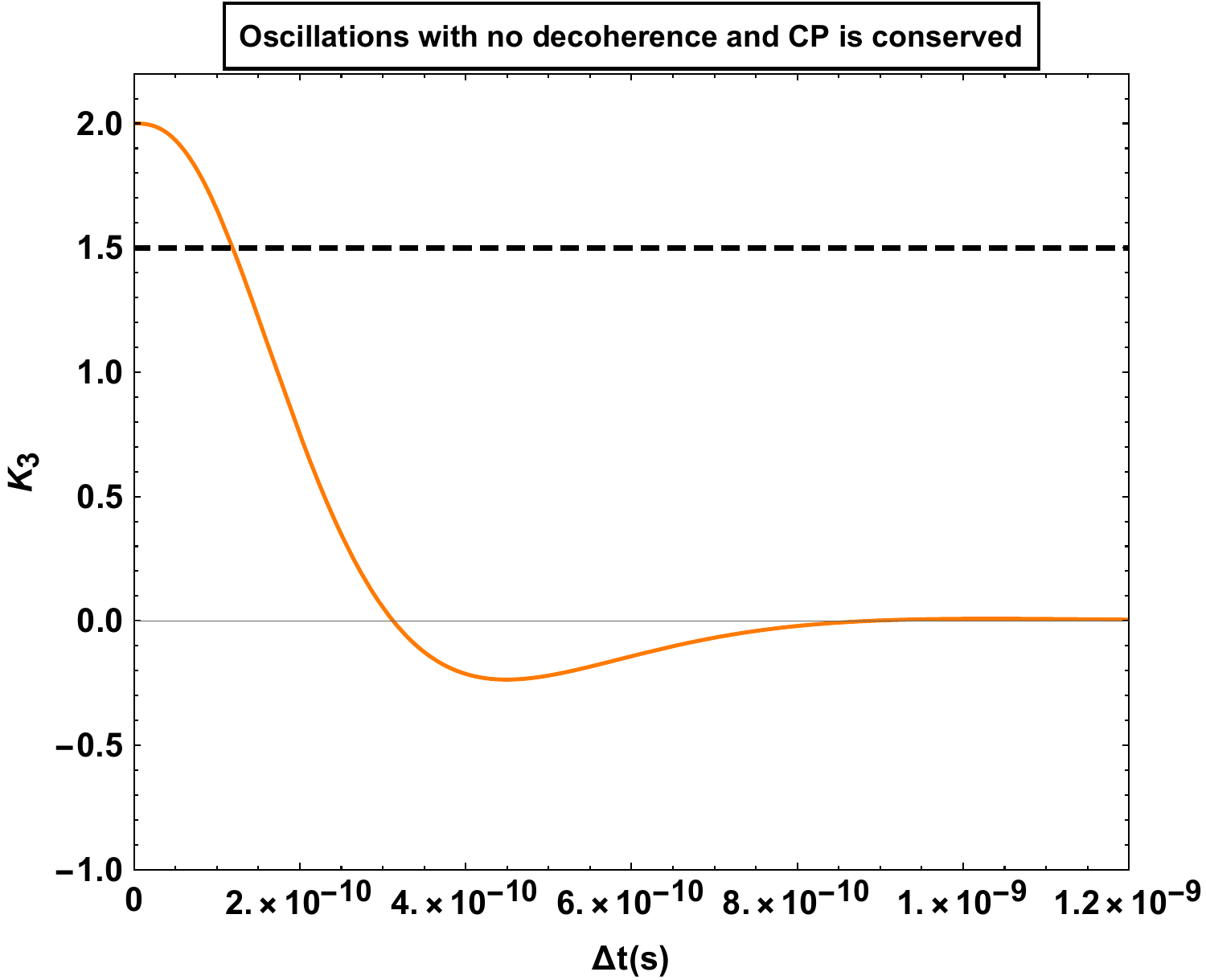}
    \caption{Plot for $K_3$ vs $\Delta t$ for CP eigenstates of $K$-mesons. The horizontal shaded line denotes the maximum quantum mechanical violation for $K_3$.}
    \label{fig:my_label}
\end{figure}
In the context of oscillations of CP, eigenstates of $K$-mesons have even more weird results to show. The value of $K_3$ for some intervals of $\Delta t$ crosses over the maximum quantum mechanical correlated value of 1.5. We can see for some initial small interval of $\Delta t$, it takes the value of $K_3=2$, much like the non-CP states; as we saw earlier, the behavior of $K_3$ over various successive intervals of $\Delta t$ remains pretty much the same, except it takes different values for $K_3$. From the perspective of LGI, the value of K crossing over the mark of the temporal Tsirelson bound has something more to hint that the measurement leads to a more significant wave function collapse than the projective measurements we saw earlier. Other experiments have shown these enhanced violations in LGI, like at $PT$ symmetry breaking point \cite{Karthik:2019wbp,Wang:17}. This could be signifying that the natural systems of mesons are never CP conserved, so this could state that they are unphysical phenomena that are limited to theoretical analysis in the context of LGI. A similar result will also follow for the $B$-meson oscillation system.

The test for LGI violation at the experiment level can be probed by precisely measuring the oscillation probabilities of neutral mesons. The neutral meson's flavor states are often examined using the tagging method. For instance, the semileptonic decay of $B^0_d$ meson results in a positron, neutrino, and a hadron. But, for $\overline{B}^0_d$ meson, the final state will have an electron. Therefore, in semileptonic decays of a neutral meson, the flavor of the meson at the time of decay is generally determined by the charge of the final state lepton. The oscillation probability of the tagged meson is then determined by identifying its final flavor-specific state~\cite{Naikoo:2018vug}.
Similarly, one can also study the regeneration processes to analyze the flavor states experimentally. For example, in $K$ meson decays, $K_L$ mesons are obtained more than the short-lived $K_S$. Letting $K_L$ shoot through a block of matter again leads to a combination of $K^0$ and $\bar{K}^0$ states. These components show dissimilar behavior of interaction within the matter. With this, one can distinctly identify where $K^0$ particle undergoes quasi-elastic scattering interactions with nucleons, and on the other hand, $\bar{K}^0$ has the potential to yield hyperons~\cite{Pais:1955sm}.

\section{Conclusion}
The study of LGI has revealed productive outcomes in deciphering the actual behavior of physical systems. Just like how Bell inequality proved to test realism based on the locality argument, similarly here we're able to test real characteristics of a system using LGI using the non-invasive measurability argument. Implementing LGI in our understanding of the meson oscillation system has equally been very useful in knowing various intrinsic aspects governing the role of oscillations. We found how these oscillation systems can be treated as exemplary open quantum systems where we saw how the interaction of the environment with the system plays out the role of decoherence \cite{Shafaq:2021lju}. LGI further helped us know how much maximum correlation exists between the oscillating states for both decay and decoherence. Decoherence generally suppresses the LGI violation compared to the situation when oscillations take place coherently. More vividly, we emphasized how Temporal \textit{Tsirelson bound} is nearly or fully achieved only whenever decay effects are neglected, i.e., taking measurements at small intervals comparable to their mean proper lifetime. There on, we also showed the fact that decay and decoherence time have different physical origins. The persisting oscillation behavior obtained in a few cases can be used to carry out a loophole-free interferometric test at the experimental level, as performed out in ref. \cite{PRXQuantum.3.010307} and the references therein using heralded single photons. Moreover, the LGI violation in neutrino oscillation has also been looked at the ongoing accelerator experiment MINOS~\cite{PhysRevLett.117.050402} and reactor experiment Daya-Bay~\cite{Fu:2017hky}. In analogy, our analysis suggests that for the experimental confirmation of LGI violations in mesonic system, the accelerator experiments should focus on the short baselines providing the source of kaons and pions close to the region of oscillation probability maxima or minima.

Recently, in the works \cite{Blasone:2024jur, Blasone:2022iwf}, it has been pointed out that unlike Bell inequalities, which provide both necessary and sufficient conditions for local realism, LGI is not an adequate requirement for fulfilling macrorealism. It is observed that under certain conditions, the optimal forms of LGI and Wigner’s form of LGI can be satisfied, hiding the quantum nature of the physical system. There are additional macrorealistic inequalities, particularly no-signaling in time and the Arrow of Time, which can provide more robust compatibility to test macrorealism. In \cite{Blasone:2021mbc}, the authors also discuss the comparison of LGI violation in quantum mechanics with quantum field theory. The temporal violation is much more robust in the QFT framework than in quantum mechanics. The temporal analog for Tsirelson’s bound is studied using the flavor mass uncertainty relation, which sets an upper bound for LGI violations at all energy scales. Depending on the measurement settings, there are different variants of LGI and Wigner forms of LGI (WLGIs), which help study the meson and neutrino subatomic systems.

 Finally, we also produced the naive treatment of meson oscillation in terms of CP eigenstates, which showed weird results of LGI violation, where the violations were over-enhanced (or crossed the mark of Tsirelson bound). This could signify some hidden phenomenon playing out, or it could be a mere unphysical phenomenon where CP is supposed to be conserved in the weak interaction, and mesons decay via the weak interaction. Thus, LGI helps us to unfold various aspects behind the $K$- and $B-$ meson physics at its intricate features.
 
 \section{Acknowledgement}
 
One of the authors, KS, wants to thank Prof. Prasanta K Panigrahi for the hospitality and valuable discussions during her research visit in April 2022 at IISER, Kolkata. KS is also thankful to the Ministry of Education, Govt of India, for financial support. Also, Aryabrat Mahapatra, a second-year student of MSc from IIT Bhubaneswar, would like to acknowledge Dr. Sudhanwa Patra for the fruitful discussion during the internship period of June-July 2022.
\vspace*{1.5cm}
\begin{widetext}
\begin{appendix}
\section{Evolution of density matrix in time using Kraus representation}\label{append:1}
The evolution of density  matrix for the oscillations of $K^0-\bar{K^0}$ when the initial state is  $\ket{K^0}$;
\begin{eqnarray}
 \rho_{K^0}(t) &=& \sum_i\mathcal{K}_i \rho_{K^0}(0)\mathcal{K}_i^{\dagger}\nonumber\\
                &=& \mathcal{K}_0 \rho_{K^0}(0)\mathcal{K}_0^{\dagger}+\mathcal{K}_1 \rho_{K^0}(0)\mathcal{K}_1^{\dagger}+...+\mathcal{K}_5 \rho_{K^0}(0)\mathcal{K}_5^{\dagger}\nonumber
\end{eqnarray}
where $\rho_{K^0}(0)=\ket{K^0}\bra{K^0}$, \hspace{0.24cm} i.e.,\hspace{0.24cm}$\rho_{K^0}(0)=
\begin{pmatrix}
1\\
0\\
0
\end{pmatrix}
\begin{pmatrix}
1 & 0 & 0
\end{pmatrix}=
\begin{pmatrix}
1 & 0 & 0\\
0 & 0 & 0\\
0 & 0 & 0
\end{pmatrix}.$ 
So,

 \begin{eqnarray}
  \mathcal{K}_0 \rho_{K^0}(0)\mathcal{K}_0^{\dagger} &=& (\ket{0}\bra{0})(\ket{K^0}\bra{K^0})(\ket{0}\bra{0})^{\dagger}=0\nonumber\\
  \mathcal{K}_1 \rho_{K^0}(0)\mathcal{K}_1^{\dagger} &=& (\mathcal{C}_{1+}(\ket{K^0}\bra{K^0}+\ket{\bar{K^0}}\bra{\bar{K^0}}+\mathcal{C}_{1-}(\frac{1+\epsilon}{1-\epsilon}\ket{K^0}\bra{\bar{K^0}}+\frac{1-\epsilon}{1+\epsilon}\ket{\bar{K^0}}\bra{K^0}))(\ket{K^0}\bra{K^0})\nonumber\\
  &&
  (\mathcal{C}_{1+}(\ket{K^0}\bra{K^0}+\ket{\bar{K^0}}\bra{\bar{K^0}}+\mathcal{C}_{1-}(\frac{1+\epsilon}{1-\epsilon}\ket{K^0}\bra{\bar{K^0}}+\frac{1-\epsilon}{1+\epsilon}\ket{\bar{K^0}}\bra{K^0}))^{\dagger}\nonumber\\
  &=& |\mathcal{C}_{1+}|^2\ket{K^0}\bra{K^0}+\mathcal{C}_{1+}\mathcal{C}_{1-}^*(\frac{1-\epsilon}{1+\epsilon})^*\ket{K^0}\bra{\bar{K^0}}+\mathcal{C}_{1+}^*\mathcal{C}_{1-}(\frac{1-\epsilon}{1+\epsilon})\ket{\bar{K^0}}\bra{K^0}+|\mathcal{C}_{1-}|^2|\frac{1-\epsilon}{1+\epsilon}|^2\ket{\bar{K^0}}\bra{\bar{K^0}}\nonumber\\
  &=&
  \begin{pmatrix}
  |\mathcal{C}_{1+}|^2 & \mathcal{C}_{1+}\mathcal{C}_{1-}^*(\frac{1-\epsilon}{1+\epsilon})^* & 0 \\
  \mathcal{C}_{1+}^*\mathcal{C}_{1-}(\frac{1-\epsilon}{1+\epsilon}) & |\mathcal{C}_{1-}|^2|\frac{1-\epsilon}{1+\epsilon}|^2 & 0 \\
  0 & 0 & 0 \\
  \end{pmatrix}\nonumber\\
 \mathcal{K}_2 \rho_{K^0}(0)\mathcal{K}_2^{\dagger} &=& (\mathcal{C}_{2}(\frac{1}{1+\epsilon}\ket{0}\bra{K^0}+\frac{1}{1-\epsilon}\ket{0}\bra{\bar{K^0}}))(\ket{K^0}\bra{K^0})(\mathcal{C}_{2}(\frac{1}{1+\epsilon}\ket{0}\bra{K^0}+\frac{1}{1-\epsilon}\ket{0}\bra{\bar{K^0}}))^{\dagger}\nonumber\\
 &=& |\mathcal{C}_2|^2\left|\frac{1}{1+\epsilon}\right|^2\ket{0}\bra{0}\nonumber\\
 &=& 
 \begin{pmatrix}
 0 & 0 & 0\\
 0 & 0 & 0\\
 0 & 0 & |\mathcal{C}_2|^2\left|\frac{1}{1+\epsilon}\right|^2\\
 \end{pmatrix}\nonumber\\
   \mathcal{K}_3 \rho_{K^0}(0)\mathcal{K}_3^{\dagger} &=& (\mathcal{C}_{3+}\frac{1}{1+\epsilon}\ket{0}\bra{K^0}+\mathcal{C}_{3-}\frac{1}{1-\epsilon}\ket{0}\bra{\bar{K^0}})(\ket{K^0}\bra{K^0})(\mathcal{C}_{3+}\frac{1}{1+\epsilon}\ket{0}\bra{K^0}+\mathcal{C}_{3-}\frac{1}{1-\epsilon}\ket{0}\bra{\bar{K^0}})^{\dagger}\nonumber\\
   &=& |\mathcal{C}_{3+}|^2\left|\frac{1}{1+\epsilon}\right|^2\ket{0}\bra{0}\nonumber\\
   &=&
   \begin{pmatrix}
    0 & 0 & 0\\
 0 & 0 & 0\\
 0 & 0 & |\mathcal{C}_{3+}|^2\left|\frac{1}{1+\epsilon}\right|^2\\
   \end{pmatrix}\nonumber\\\mathcal{K}_4 \rho_{K^0}(0)\mathcal{K}_4^{\dagger} &=&  (\mathcal{C}_{4}(\ket{K^0}\bra{K^0}+\ket{\bar{K^0}}\bra{\bar{K^0}}+\frac{1+\epsilon}{1-\epsilon}\ket{K^0}\bra{\bar{K^0}}+\frac{1-\epsilon}{1+\epsilon}\ket{\bar{K^0}}\bra{K^0}))(\ket{K^0}\bra{K^0})\nonumber\\
    &&
    (\mathcal{C}_{4}(\ket{K^0}\bra{K^0}+\ket{\bar{K^0}}\bra{\bar{K^0}}+\frac{1+\epsilon}{1-\epsilon}\ket{K^0}\bra{\bar{K^0}}+\frac{1-\epsilon}{1+\epsilon}\ket{\bar{K^0}}\bra{K^0}))^{\dagger}\nonumber\\
    &=& |\mathcal{C}_4|^2\ket{K^0}\bra{K^0}+|\mathcal{C}_4|^2(\frac{1-\epsilon}{1+\epsilon})^*\ket{K^0}\bra{\bar{K^0}}+|\mathcal{C}_4|^2\left(\frac{1-\epsilon}{1+\epsilon}\right)^*\ket{\bar{K^0}}\bra{K^0}+|\mathcal{C}_4|^2\left|\frac{1-\epsilon}{1+\epsilon}\right|^2\ket{\bar{K^0}}\bra{\bar{K^0}}\nonumber\\
    &=&
    \begin{pmatrix}
    |\mathcal{C}_4|^2   & |\mathcal{C}_4|^2(\frac{1-\epsilon}{1+\epsilon})^* & 0\\
 |\mathcal{C}_4|^2(\frac{1-\epsilon}{1+\epsilon}) & |\mathcal{C}_4|^2\left|\frac{1-\epsilon}{1+\epsilon}\right|^2 & 0\\
 0 & 0 & 0
    \end{pmatrix}\nonumber\\
     \mathcal{K}_5 \rho_{K^0}(0)\mathcal{K}_5^{\dagger} &=& (\mathcal{C}_{5}(\ket{K^0}\bra{K^0}+\ket{\bar{K^0}}\bra{\bar{K^0}} -\frac{1+\epsilon}{1-\epsilon}\ket{K^0}\bra{\bar{K^0}}-\frac{1-\epsilon}{1+\epsilon}\ket{\bar{K^0}}\bra{K^0}))(\ket{K^0}\bra{K^0})\nonumber\\
     &&
     (\mathcal{C}_{5}(\ket{K^0}\bra{K^0}+\ket{\bar{K^0}}\bra{\bar{K^0}} -\frac{1+\epsilon}{1-\epsilon}\ket{K^0}\bra{\bar{K^0}}-\frac{1-\epsilon}{1+\epsilon}\ket{\bar{K^0}}\bra{K^0}))^{\dagger}\nonumber\\
     &=&  |\mathcal{C}_5|^2\ket{K^0}\bra{K^0}-|\mathcal{C}_5|^2(\frac{1-\epsilon}{1+\epsilon})^*\ket{K^0}\bra{\bar{K^0}}-|\mathcal{C}_5|^2\left(\frac{1-\epsilon}{1+\epsilon}\right)\ket{\bar{K^0}}\bra{K^0}+|\mathcal{C}_5|^2\left|\frac{1-\epsilon}{1+\epsilon}\right|^2\ket{\bar{K^0}}\bra{\bar{K^0}}\nonumber\\
     &=&
     \begin{pmatrix}
       |\mathcal{C}_5|^2 & -|\mathcal{C}_5|^2(\frac{1-\epsilon}{1+\epsilon})^* & 0\\
 -|\mathcal{C}_5|^2(\frac{1-\epsilon}{1+\epsilon}) & |\mathcal{C}_5|^2\left|\frac{1-\epsilon}{1+\epsilon}\right|^2 & 0\\
 0 & 0 & 0
     \end{pmatrix}\nonumber
   \end{eqnarray}

\begin{eqnarray}
   \therefore  \rho_{K^0}(t)=
   \begin{pmatrix}
  |\mathcal{C}_{1+}|^2 +|\mathcal{C}_4|^2 +|\mathcal{C}_5|^2 & (\frac{1-\epsilon}{1+\epsilon})^*(\mathcal{C}_{1+}\mathcal{C}_{1-}^*+|\mathcal{C}_4|^2-|\mathcal{C}_5|^2) & 0 \\
   (\frac{1-\epsilon}{1+\epsilon})(\mathcal{C}_{1+}\mathcal{C}_{1-}^*+|\mathcal{C}_4|^2-|\mathcal{C}_5|^2)  & \left|\frac{1-\epsilon}{1+\epsilon}\right|^2(|\mathcal{C}_{1-}|^2 +|\mathcal{C}_4|^2 +|\mathcal{C}_5|^2)  & 0 \\
    0  & 0 & \left|\frac{1}{1+\epsilon}\right|^2 (|\mathcal{C}_{2}|^2+|\mathcal{C}_{3+}|^2)
   \end{pmatrix}\nonumber
\end{eqnarray}

Now, using the coefficient values from equations (\ref{eqn-36}), we will get the desired matrix as given in equation (\ref{eqn-39}).

\section{Computing the two-time correlation function in terms of Kraus operators}\label{append:2}
The general expression for probabilities as obtained earlier is:
\begin{eqnarray}
   p(^at_i)q(^bt_j|^at_i)= Tr\left\{\Pi^b\sum_{\nu,\mu}\mathcal{K}_{\nu}(t_j-t_i)\Pi^a\sum_{\mu}\mathcal{K}_{\mu}(t_i)\rho(0)\mathcal{K}_{\mu}^{\dagger}(t_i)\Pi^a\mathcal{K}_{\nu}^{\dagger}(t_j-t_i)
\right\}\nonumber
\end{eqnarray}
The projector operators for the dichotomic observable which are involved here also satisfies a relation as $\Pi^++\Pi^-=I$ 
The two-time correlation function is expanded in these probabilities as \cite{PhysRevA.88.022104}:
\begin{eqnarray}
   C(t_i,t_j) = p(^+t_i)q(^+t_j|^+t_i)-p(^+t_i)q(^-t_j|^+t_i)-p(^-t_i)q(^+t_j|^-t_i)+p(^-t_i)q(^-t_j|^-t_i)\nonumber
\end{eqnarray}

where,

\begin{eqnarray}
p(^+t_i)q(^+t_j|^+t_i) &=& Tr\left\{\Pi^+\sum_{\nu,\mu}\mathcal{K}_{\nu}(t_j-t_i)\Pi^+\mathcal{K}_{\mu}(t_i)\rho(0)\mathcal{K}_{\mu}^{\dagger}(t_i)\Pi^+\mathcal{K}_{\nu}^{\dagger}(t_j-t_i)
\right\}\nonumber\\
    p(^-t_i)q(^-t_j|^-t_i) &=& Tr\left\{\Pi^-\sum_{\nu,\mu}\mathcal{K}_{\nu}(t_j-t_i)\Pi^-\mathcal{K}_{\mu}(t_i)\rho(0)\mathcal{K}_{\mu}^{\dagger}(t_i)\Pi^-\mathcal{K}_{\nu}^{\dagger}(t_j-t_i)
\right\}\nonumber\\
&=& Tr\left\{(I-\Pi^+)\sum_{\nu,\mu}\mathcal{K}_{\nu}(t_j-t_i)(I-\Pi^+)\mathcal{K}_{\mu}(t_i)\rho(0)\mathcal{K}_{\mu}^{\dagger}(t_i)(I-\Pi^+)\mathcal{K}_{\nu}^{\dagger}(t_j-t_i)
\right\}\nonumber\\
&=& Tr\left\{\sum_{\nu,\mu}\mathcal{K}_{\nu}(t_j-t_i)(I-\Pi^+)\mathcal{K}_{\mu}(t_i)\rho(0)\mathcal{K}_{\mu}^{\dagger}(t_i)(I-\Pi^+)\mathcal{K}_{\nu}^{\dagger}(t_j-t_i)
\right\}\nonumber\\
&&
-Tr\left\{\Pi^+\sum_{\nu,\mu}\mathcal{K}_{\nu}(t_j-t_i)(I-\Pi^+)\mathcal{K}_{\mu}(t_i)\rho(0)\mathcal{K}_{\mu}^{\dagger}(t_i)(I-\Pi^+)\mathcal{K}_{\nu}^{\dagger}(t_j-t_i) \right\}\nonumber\\
&=&
 \resizebox{.86\hsize}{!}{$Tr\left\{\sum_{\nu,\mu}\mathcal{K}_{\nu}(t_j-t_i)\mathcal{K}_{\mu}(t_i)\rho(0)\mathcal{K}_{\mu}^{\dagger}(t_i)\mathcal{K}_{\nu}^{\dagger}(t_j-t_i)
\right\}-Tr\left\{\sum_{\nu,\mu}\mathcal{K}_{\nu}(t_j-t_i)\mathcal{K}_{\mu}(t_i)\rho(0)\mathcal{K}_{\mu}^{\dagger}(t_i)\Pi^+\mathcal{K}_{\nu}^{\dagger}(t_j-t_i)
\right\}$}\nonumber\\
&&
\resizebox{.86\hsize}{!}{$-Tr\left\{\sum_{\nu,\mu}\mathcal{K}_{\nu}(t_j-t_i)\Pi^+\mathcal{K}_{\mu}(t_i)\rho(0)\mathcal{K}_{\mu}^{\dagger}(t_i)\mathcal{K}_{\nu}^{\dagger}(t_j-t_i)
\right\}+Tr\left\{\sum_{\nu,\mu}\mathcal{K}_{\nu}(t_j-t_i)\Pi^+\mathcal{K}_{\mu}(t_i)\rho(0)\mathcal{K}_{\mu}^{\dagger}(t_i)\Pi^+\mathcal{K}_{\nu}^{\dagger}(t_j-t_i)
\right\}$}\nonumber\\
&&
\resizebox{.86\hsize}{!}{$-Tr\left\{\Pi^+\sum_{\nu,\mu}\mathcal{K}_{\nu}(t_j-t_i)\mathcal{K}_{\mu}(t_i)\rho(0)\mathcal{K}_{\mu}^{\dagger}(t_i)\mathcal{K}_{\nu}^{\dagger}(t_j-t_i)
\right\}+Tr\left\{\Pi^+\sum_{\nu,\mu}\mathcal{K}_{\nu}(t_j-t_i)\mathcal{K}_{\mu}(t_i)\rho(0)\mathcal{K}_{\mu}^{\dagger}(t_i)\Pi^+\mathcal{K}_{\nu}^{\dagger}(t_j-t_i)
\right\}$}\nonumber\\
&&
\resizebox{.86\hsize}{!}{$+Tr\left\{\Pi^+\sum_{\nu,\mu}\mathcal{K}_{\nu}(t_j-t_i)\Pi^+\mathcal{K}_{\mu}(t_i)\rho(0)\mathcal{K}_{\mu}^{\dagger}(t_i)\mathcal{K}_{\nu}^{\dagger}(t_j-t_i)
\right\}-Tr\left\{\Pi^+\sum_{\nu,\mu}\mathcal{K}_{\nu}(t_j-t_i)\Pi^+\mathcal{K}_{\mu}(t_i)\rho(0)\mathcal{K}_{\mu}^{\dagger}(t_i)\Pi^+\mathcal{K}_{\nu}^{\dagger}(t_j-t_i)
\right\}$}\nonumber\\
p(^-t_i)q(^+t_j|^-t_i) &=& Tr\left\{\Pi^+\sum_{\nu,\mu}\mathcal{K}_{\nu}(t_j-t_i)\Pi^-\mathcal{K}_{\mu}(t_i)\rho(0)\mathcal{K}_{\mu}^{\dagger}(t_i)\Pi^-\mathcal{K}_{\nu}^{\dagger}(t_j-t_i)
\right\}\nonumber\\
&=& Tr\left\{\Pi^+\sum_{\nu,\mu}\mathcal{K}_{\nu}(t_j-t_i)(I-\Pi^+)\mathcal{K}_{\mu}(t_i)\rho(0)\mathcal{K}_{\mu}^{\dagger}(t_i)(I-\Pi^+)\mathcal{K}_{\nu}^{\dagger}(t_j-t_i)
\right\}\nonumber\\
&=&
\resizebox{.86\hsize}{!}{$Tr\left\{\Pi^+\sum_{\nu,\mu}\mathcal{K}_{\nu}(t_j-t_i)\mathcal{K}_{\mu}(t_i)\rho(0)\mathcal{K}_{\mu}^{\dagger}(t_i)\mathcal{K}_{\nu}^{\dagger}(t_j-t_i)
\right\}-Tr\left\{\Pi^+\sum_{\nu,\mu}\mathcal{K}_{\nu}(t_j-t_i)\mathcal{K}_{\mu}(t_i)\rho(0)\mathcal{K}_{\mu}^{\dagger}(t_i)\Pi^+\mathcal{K}_{\nu}^{\dagger}(t_j-t_i)
\right\}$}\nonumber\\
&&
\resizebox{.86\hsize}{!}{$-Tr\left\{\Pi^+\sum_{\nu,\mu}\mathcal{K}_{\nu}(t_j-t_i)\Pi^+\mathcal{K}_{\mu}(t_i)\rho(0)\mathcal{K}_{\mu}^{\dagger}(t_i)\mathcal{K}_{\nu}^{\dagger}(t_j-t_i)
\right\}+Tr\left\{\Pi^+\sum_{\nu,\mu}\mathcal{K}_{\nu}(t_j-t_i)\Pi^+\mathcal{K}_{\mu}(t_i)\rho(0)\mathcal{K}_{\mu}^{\dagger}(t_i)\Pi^+\mathcal{K}_{\nu}^{\dagger}(t_j-t_i)
\right\}$}\nonumber\\
p(^+t_i)q(^-t_j|^+t_i) &=& Tr\left\{\Pi^-\sum_{\nu,\mu}\mathcal{K}_{\nu}(t_j-t_i)\Pi^+\mathcal{K}_{\mu}(t_i)\rho(0)\mathcal{K}_{\mu}^{\dagger}(t_i)\Pi^+\mathcal{K}_{\nu}^{\dagger}(t_j-t_i)
\right\}\nonumber\\
&=& Tr\left\{(I-\Pi^+)\sum_{\nu,\mu}\mathcal{K}_{\nu}(t_j-t_i)\Pi^+\mathcal{K}_{\mu}(t_i)\rho(0)\mathcal{K}_{\mu}^{\dagger}(t_i)\Pi^+\mathcal{K}_{\nu}^{\dagger}(t_j-t_i)
\right\}\nonumber\\
&=&
\resizebox{.86\hsize}{!}{$Tr\left\{\sum_{\nu,\mu}\mathcal{K}_{\nu}(t_j-t_i)\mathcal{K}_{\mu}(t_i)\rho(0)\mathcal{K}_{\mu}^{\dagger}(t_i)\mathcal{K}_{\nu}^{\dagger}(t_j-t_i)
\right\}-Tr\left\{\Pi^+\sum_{\nu,\mu}\mathcal{K}_{\nu}(t_j-t_i)\Pi^+\mathcal{K}_{\mu}(t_i)\rho(0)\mathcal{K}_{\mu}^{\dagger}(t_i)\Pi^+\mathcal{K}_{\nu}^{\dagger}(t_j-t_i)
\right\}$}\nonumber
\end{eqnarray}

Thus, correlation function will now become:
\begin{eqnarray}
C(t_i,t_j) &=& 
\resizebox{.86\hsize}{!}{$1-2Tr\left\{\Pi^+\sum_{\nu,\mu}\mathcal{K}_{\nu}(t_j-t_i)\mathcal{K}_{\mu}(t_i)\rho(0)\mathcal{K}_{\mu}^{\dagger}(t_i)\mathcal{K}_{\nu}^{\dagger}(t_j-t_i)
\right\}-Tr\left\{\sum_{\nu,\mu}\mathcal{K}_{\nu}(t_j-t_i)\Pi^+\mathcal{K}_{\mu}(t_i)\rho(0)\mathcal{K}_{\mu}^{\dagger}(t_i)\mathcal{K}_{\nu}^{\dagger}(t_j-t_i)
\right\}$}\nonumber\\
&&
\resizebox{.86\hsize}{!}{$-Tr\left\{\sum_{\nu,\mu}\mathcal{K}_{\nu}(t_j-t_i)\mathcal{K}_{\mu}(t_i)\rho(0)\mathcal{K}_{\mu}^{\dagger}(t_i)\Pi^+\mathcal{K}_{\nu}^{\dagger}(t_j-t_i)
\right\}+2Tr\left\{\Pi^+\sum_{\nu,\mu}\mathcal{K}_{\nu}(t_j-t_i)\Pi^+\mathcal{K}_{\mu}(t_i)\rho(0)\mathcal{K}_{\mu}^{\dagger}(t_i)\mathcal{K}_{\nu}^{\dagger}(t_j-t_i)
\right\}$}\nonumber\\
&&
\resizebox{.43\hsize}{!}{$+Tr\left\{\Pi^+\sum_{\nu,\mu}\mathcal{K}_{\nu}(t_j-t_i)\mathcal{K}_{\mu}(t_i)\rho(0)\mathcal{K}_{\mu}^{\dagger}(t_i)\Pi^+\mathcal{K}_{\nu}^{\dagger}(t_j-t_i)
\right\}$}\nonumber
\end{eqnarray}

The second term gives the probability of getting $+1$ when $\rho(0)$ is evolved till $t_j$ and third and fourth terms gives the probability  of getting $+1$ when $\rho(0)$ is evolved till $t_i$,i.e. as following:
\begin{eqnarray}
p(^+t_i) &=& Tr\left\{ \sum_{\mu}\Pi^+\mathcal{K}_{\mu}(t_i)\rho(0)\mathcal{K}_{\mu}^{\dagger}(t_i)\right\}\nonumber\\
&=& Tr\left\{\sum_{\nu,\mu}\mathcal{K}_{\nu}(t_j-t_i)\left[\Pi^+\mathcal{K}_{\mu}(t_i)\rho(0)\mathcal{K}_{\mu}^{\dagger}(t_i)\right]\mathcal{K}_{\nu}^{\dagger}(t_j-t_i)
\right\}\nonumber
\end{eqnarray}

The last two terms can be further rewritten as $2g(t_i,t_j)$ and its hermitian conjugate, where,
\begin{eqnarray}
g(t_i,t_j)=Tr\left\{\rho(0)\sum_{\nu,\mu}\mathcal{K}^{\dagger}_{\mu}(t_i)\mathcal{K}^{\dagger}_\nu(t_j-t_i)\Pi^+\mathcal{K}_\nu(t_j-t_i)\mathcal{K}_{\mu}(t_i) \right\}\nonumber
\end{eqnarray}
or 
\begin{eqnarray}
g(t_i,t_j) &=& Tr\left\{\sum_{\nu,\mu}\Pi^+\mathcal{K}_\nu(t_j-t_i)\Pi^+ \mathcal{K}_{\mu}(t_i)\rho(0){K}^{\dagger}_{\mu}(t_i)\mathcal{K}^{\dagger}_\nu(t_j-t_i)\right\}\nonumber\\
&=& Tr\left\{\Pi^+\sum_{\nu}\mathcal{K}_\nu(t_j-t_i)\rho(t_i) \mathcal{K}^{\dagger}_\nu(t_j-t_i)\right\}\nonumber
\end{eqnarray}
Hence, two-time correlation can be written as in equation (\ref{eqn-45}), i.e.;
\begin{eqnarray}
 C(t_i,t_j)=1-2p(^+t_i)-2p(^+t_j)+4Re[g(t_i,t_j)]\nonumber
\end{eqnarray}

 \end{appendix}
\vspace*{1cm}

\section*{Data Availability Statement} No Data associated in the manuscript.
\end{widetext}

\bibliographystyle{utcaps_mod}
\bibliography{lgi.bib}

\end{document}